\documentclass{article}
\usepackage{graphicx} 
\usepackage{amsmath}
\usepackage{jheppub}
\usepackage{adjustbox}
\usepackage{physics}
\usepackage{caption}
\usepackage{tikz}
\usepackage{tikz-cd}
\usepackage{parskip}
\usepackage[normalem]{ulem}
\usepackage{setspace}
\usepackage{cleveref}
\usepackage{orcidlink}
\usepackage{float}

\usepackage{listings}
\usepackage[dvipsnames]{xcolor}
\usepackage{mdframed}

\usepackage{cleveref}

\definecolor{light-gray}{gray}{0.95}

\definecolor{gh-fg}{HTML}{000000}
\definecolor{gh-comment}{HTML}{007ACC}
\definecolor{gh-keyword}{HTML}{E60000}
\definecolor{gh-func}{HTML}{FF7F00}
\definecolor{gh-string}{HTML}{999999} 
\definecolor{gh-number}{HTML}{000000}
\definecolor{gh-linenum}{HTML}{666666}
\definecolor{gh-frame}{HTML}{BBBBBB}

\lstnewenvironment{fullpagecode}[1][]%
  {\noindent\begin{minipage}{\textwidth}\begin{lstlisting}[#1]}%
  {\end{lstlisting}\end{minipage}}
\lstdefinestyle{githubMathematica}{
  language=Mathematica,
  backgroundcolor=\color{light-gray},
  basicstyle={\small\setstretch{1}\def\fvm@Scale{0.85}\ttfamily\selectfont\color{gh-fg}},
  commentstyle=\itshape\color{gh-comment},
  keywordstyle=\bfseries\color{gh-fg},
  stringstyle=\color{gh-string},
  identifierstyle=\color{gh-fg},
  numbers=right,
  numberstyle={\scriptsize\sffamily\color{gh-linenum}},
  numbersep=8pt,
  frame=lines,
  rulecolor=\color{gh-frame},
  tabsize=4,
  showstringspaces=false,
  breaklines=false,
  numbers=none,
  columns=fixed,
  basewidth=0.5em,
  aboveskip=1em,
  belowskip=0pt,
  emph={FindIrreducibleMonomials,BuildPolynomialSystem,
        ReconstructPolynomialRemainder,BuildCompanionMatrices,
        BuildTargetCompanionMatrices,ReconstructTargetCompanionMatrices,
        BuildCharacteristicPolynomials,ReconstructCharacteristicPolynomials,
        FFDet,SortVariables,FindEliminationMonomials,BuildEliminationSystems,ReconstructEliminationSystems,MagicQ,FindMagicRelations,FFPrimeNo,CATSyz,j},
  emphstyle=\bfseries\color{purple},
  literate=
    *{0}{{{\color{gh-number}0}}}1 {1}{{{\color{gh-number}1}}}1
     {2}{{{\color{gh-number}2}}}1 {3}{{{\color{gh-number}3}}}1
     {4}{{{\color{gh-number}4}}}1 {5}{{{\color{gh-number}5}}}1
     {6}{{{\color{gh-number}6}}}1 {7}{{{\color{gh-number}7}}}1
     {8}{{{\color{gh-number}8}}}1 {9}{{{\color{gh-number}9}}}1
     {.0}{{{\color{gh-number}.0}}}2 {.1}{{{\color{gh-number}.1}}}2
     {.2}{{{\color{gh-number}.2}}}2 {.3}{{{\color{gh-number}.3}}}2
     {.4}{{{\color{gh-number}.4}}}2 {.5}{{{\color{gh-number}.5}}}2
     {.6}{{{\color{gh-number}.6}}}2 {.7}{{{\color{gh-number}.7}}}2
     {.8}{{{\color{gh-number}.8}}}2 {.9}{{{\color{gh-number}.9}}}2
}
\usepackage{fontawesome5}



\usepackage[most]{tcolorbox}
\tcbset{
    enhanced jigsaw, 
    colback=gray!5!white,
    colframe=gray!20!white,
    coltitle=black,
    fonttitle=\bfseries
}
\newtcolorbox[auto counter,number within=section]{mybox}[2][]{%
}
\newtcolorbox[auto counter]{myboxtitle}[2][]{%
fonttitle=\bfseries
,title=#2
}

\usetikzlibrary{shapes.geometric, arrows}

\newcommand{\id}{\mathrm{d}}

\newcommand{\crit}{\mathrm{Crit}}

\title{Magic Relations and Critical Varieties of Feynman Integrals}

\author[a]{Giulio Crisanti,\orcidlink{0009-0009-3053-2394}}
\author[b]{Hjalte Frellesvig,\orcidlink{0000-0002-7367-4861}}
\author[c]{Andrzej Pokraka,\orcidlink{0000-0003-1186-4624}}
\author[a,d,e]{Sid Smith,\orcidlink{0009-0007-7799-0136}}

\affiliation[a]{Higgs Centre for Theoretical Physics, University of Edinburgh, James Clerk Maxwell Building, Peter Guthrie Tait Road, Edinburgh, EH9 3FD, United Kingdom}
\affiliation[b]{Zhejiang Institute of Modern Physics, School of Physics, Zhejiang University, Hangzhou
310027, China}
\affiliation[c]{Institute of Physics, University of Amsterdam, Amsterdam, 1098 XH, The Netherlands}
\affiliation[d]{Dipartimento di Fisica e Astronomia, Universita di Padova, Via Marzolo 8, 35131 Padova, Italy}
\affiliation[e]{INFN, Sezione di Padova, Via Marzolo 8, I-35131 Padova, Italy}

\emailAdd{g.crisanti@ed.ac.uk}
\emailAdd{0025056@zju.edu.cn}
\emailAdd{a.m.pokraka@uva.nl}
\emailAdd{sid.smith@ed.ac.uk}

\abstract{
Magic relations are a class of integration-by-parts identities where all integrals in the generating sector drop out. 
Since their presence causes several otherwise successful methods in the Feynman-integral computational pipeline to break down, they are important to detect and understand. 
In this paper, we take a first step toward a systematic characterization of such identities. 
Specifically, we observe and argue that the occurrence of magic relations always coincides with the presence of higher-dimensional critical varieties in the generating sector. This provides a practical computational test to check if a family of Feynman integrals can contain magic relations and a way to find them, and we implement this in the ancillary \textsc{Mathematica} file \texttt{Magic-Test.m}.
Additionally, we discuss how to count the number of master integrals in the presence of higher-dimensional critical varieties, classify the behavior of magic relations under symmetries, and we discuss their interplay with cuts.
}

\begin{document}

\maketitle

\newpage
\section{Introduction}
\nocite{Lee:2013hzt, Maierhofer:2018gpa}
Feynman integrals are ubiquitous in perturbative high energy physics. 
They are essential tools for computing scattering cross sections relevant for high precision theoretical predictions for particle colliders such as the Large Hadron Collider (LHC). More broadly, Feynman-like integrals also appear in the study of gravitational waves, inflationary cosmology, solid state physics, and beyond.
In fact, Feynman-like integrals exhibit a plethora of extremely rich mathematical structures (see e.g. ref.~\cite{Weinzierl:2022eaz} and the references therein) that are of great interest to both mathematicians and mathematical physicists.

State-of-the-art computations relevant for LHC physics involve the seemingly unfeasible task of computing $10^5$--$10^6$ Feynman integrals. 
Fortunately, this number can be reduced by many orders of magnitude through the usage of \emph{integration-by-parts} identities (IBP) \cite{Chetyrkin:1981qh,Laporta:2000dsw}.  
IBP reduction algorithms work by generating linear relations between Feynman \textit{integrands} (or more technically, elements of the twisted cohomology group \cite{Mastrolia:2018uzb}) and using them to express any Feynman integral in terms of a basis of integrals called \textit{master integrals}. Due to the vast reduction in combinatorial complexity, efficient and robust IBP algorithms are fundamental to modern state of the art scattering computations.

Nearly all IBP identities relate Feynman integrals within a sector and its subsectors. Nevertheless in some cases there exist special kinds of IBP identities, known as ``magic relations'' \cite{Maierhofer:2018gpa,Smirnov:2013dia,Lee:2010ik}. These are non trivial identities that relate integrals in subsectors 
without any integral from the generating sector appearing in the relation\footnote{%
Around 2007~\cite{Drummond:2006rz} (and to some extent still~\cite{He:2025vqt}), the phrase ``magic identities'' was used for certain integral identities now understood to be a consequence of dual conformal invariance~\cite{Drummond:2008vq}. That is unrelated to the magic relations discussed in this publication.}.

The study of magic relations is important as several powerful tools utilized to simplify IBP reductions break down in their presence. In particular, magic relations are fundamental obstructions to cut-based IBP reconstruction methods implemented in modern IBP software~\cite{Larsen:2015ped,Wu:2023upw,Guan:2024byi,Smirnov:2025prc,Lange:2025fba,Wu:2025aeg}. These identities also cause issues with
intersection theory~\cite{Mastrolia:2018uzb, Frellesvig:2019uqt} (see also~\cite{Mizera:2019ose, Weinzierl:2020gda}) in the framework of {relative twisted cohomology}~\cite{Caron-Huot:2021xqj, Caron-Huot:2021iev, Fontana:2023amt, Brunello:2023rpq, Brunello:2023fef,Brunello:2024tqf,Crisanti:2024onv, Lu:2024dsb, De:2023xue, De:2024zic, Glew:2025ypb}. Despite their relative rarity, it is thus important to know if magic relations may arise before performing integral reductions. 

With this manuscript we present a systematic study of magic relations. Our most important result is the observation of a direct connection between the appearance of magic relations and the presence of higher dimensional critical varieties relevant for counting master integrals. We also provide an argument (under certain assumptions) for the validity of this connection. 

This relation between seemingly distinct phenomena allows for an efficient and practical test for detecting (and computing) magic relations without having to generate or solve large IBP systems beforehand. To this end, two proof of concept $\textsc{Mathematica}$ functions are provided in the ancillary file \texttt{Magic-Test.m} which check for and build magic relations. 

Furthermore we provide a discussion of the Lee--Pomeransky criterion~\cite{Lee:2013hzt} for counting the number of master integrals, detailing how the prescription changes in the presence of higher dimensional critical varieties. We refer to the prescription in the absence of these higher dimensional varieties as the \textit{restricted} Lee--Pomeransky criterion, and the more general prescription as the \textit{full} Lee--Pomeransky criterion. 
These criteria are parallel to \textit{Morse--Bott theory} arguments.

The structure of our paper is as follows: In \cref{sec:background} we review: IBPs in various representations, sectors and spanning cuts, methods for computing the number of master integrals, as well as symmetry relations. In \cref{sec:magicrelations} we discuss what magic relations are, how they impact cuts and IBPs, and we discuss an instructive example. 
Then, we introduce the notion of a critical variety and state our central observation: \emph{that magic relations and higher dimensional critical varieties imply each other}. How symmetry relations interact with magic relations is also discussed.
In \cref{sec:proof}, we provide an argument for the central observation. This argument uses syzygies and IBPs in the Lee--Pomeransky representation, and exploits a deep connection to Koszul cohomology.
In \cref{sec:examples}, we provide a number of examples of Feynman integrals with magic relations and higher dimensional critical varieties. 
Lastly in \cref{sec:discussion} we discuss and conclude. The paper ends with three appendices: Appendix \ref{app:cut_calc} contains the derivation of the maximal cut of one of our example integrals, appendix \ref{app:leepomandmorse} discusses the link between the full Lee--Pomeransky criterion and Morse--Bott theory, while appendix \ref{app:trivialargument} contains a part of the argument for the central observation related to trivial syzygies.

\textit{Note added:} Shortly after the original publication of this manuscript, ref.~\cite{Wu:2026xea} appeared, which discusses an overlapping set of phenomena with regards to cuts and IBPs.

\section{Background}
\label{sec:background}

We begin by reviewing background information relevant for Feynman integral reduction. This includes IBPs in momentum-space, Baikov and Lee--Pomeransky representation, as well as the notions of sectors and spanning cuts, and how these ideas are important for tackling IBPs. 
We also introduce the restricted Lee--Pomeransky criterion for counting the number of master integrals, and briefly discuss symmetry relations.

\subsection{Integration-by-parts identities for Feynman integrals}
\label{sec:IBPs}

Integration by parts relations for Feynman integrals are most commonly discussed in the \emph{momentum representation}, which is used in most public IBP programs such as  FIRE~\cite{Smirnov:2025prc} and Kira~\cite{Lange:2025fba}. 
In this representation, $L$-loop dimensionally-regulated ($d\notin\mathbb{Z}$) Feynman integrals are integrals over rational functions of the propagators $P_i$ which are Lorentz scalars and, usually quadratic, functions of the loop-momenta ($k_{i=1,\dots,L}^\mu$) and the external momenta ($p_{i=1,\dots,N}^\mu$). In momentum representation, a generic Feynman integral is
\begin{align}\label{eq:FI}
    I_{a_1 \ldots a_n} &= \int_{\mathbb{R}^{dL}} \frac{d^d k_1}{i \pi^{d/2}} \cdots \frac{d^d k_L}{i \pi^{d/2}} \frac{P_{n_P{+}1}^{-a_{n_P+1}} \cdots P_n^{-a_n}}{P_1^{a_1} \cdots P_{n_P}^{a_{n_P}}}
    \,,
\end{align}
where normalizing by $1/i\pi^{d/2}$ cancels common factors of $\pi$ appearing in the integrated expressions.
In some cases, the propagators that appear in the physical problem, $\{P_1,\dots,P_{n_P}\}$ are not enough to span the set of Lorentz invariant scalar products $\{k_i \cdot k_j, k_i \cdot p_j\}$ and one might need to introduce additional propagators $\{P_{n_P+1},\dots,P_{n}\}$ (commonly called irreducible scalar products or ISPs). 
While all $a_i \in \mathbb{Z}$, only the physical propagators are allowed to appear in the denominator. 
Therefore, only $\{a_1, \dots, a_{n_P}\} \in \mathbb{Z}$ are allowed to be positive integers; the other $\{a_{n_P+1}, \dots, a_{n}\} \in \mathbb{Z}_{\leq0}$ must be non-positive integers. 
It is common to define a \emph{sector} by the set of $a_{i=1,\dots,n_P}>0$; we elaborate on this in \cref{sec:sectors}. 

Integration-by-parts identities originate from the fact that total derivatives vanish in dimensionally regulated integrals
\begin{align}
    0 &= \int \frac{d^d k_1}{i \pi^{d/2}} \cdots \frac{d^d k_L}{i \pi^{d/2}}
    \frac{\partial}{\partial k_j^{\mu}} 
    \frac{q^{\mu} N(k_1,\cdots,k_L)}{P_1^{a_1} \cdots P_{n_P}^{a_{n_P}}}
    \,,
\label{eq:IBP}
\end{align}
where $q^{\mu} \in \{k^{\mu}_{i=1,\dots,L},p^{\mu}_{j=1,\dots,N}\}$ is any internal or external momentum associated to the physical process, and $N(k_1,\dots,k_L)$ is a rational function of Lorentz invariant scalar products (i.e., propagators and ISPs). 
Performing the derivative yields the IBP identity 
\begin{align}
    0 &= \sum_{\vec{a}} c_{\vec{a}} I_{\vec{a}} 
\label{eq:linear}
\end{align}
where the $c_{\vec{a}}$ are functions of the kinematics ($p_i \cdot p_j$ or $m_i^2$) and the space-time dimensionality $d$.

Generally, we refer to integrals with the same set of propagators $P_{i}$ but different values of the $a_{i}$ as belonging to the same \textit{family}. IBP equations therefore provide linear relations between Feynman integrals in the same family. It has been further shown that these linear relations are a consequence of the fact that Feynman integrals of a given family belong to a finite-dimensional vector space~\cite{Smirnov:2010hn, Frellesvig:2019uqt}. This implies that by solving a system of these IBP equations, one can write any Feynman integral in the family as a linear combination of a finite set of $\nu$ integrals that form a basis on this vector space
\begin{align}
    I = \sum_{i=1}^{\nu}c_{i}J_{i}\,.
\end{align}
These integrals $J_{i}$ are commonly referred to as \textit{master integrals}. In practice this reduction is extremely important as it allows for a systematic reduction by several orders of magnitude in the number of Feynman integrals that need to be computed.

\subsubsection*{IBPs in Baikov Representation} 

In many situations, it is useful to consider Feynman integrals and IBP relations in representations other than momentum space. Here, we consider the \textit{Baikov representation}~\cite{Baikov:1996iu} (in the variant known as \textit{standard Baikov}~\cite{Frellesvig:2024ymq}). 
In this representation, a generic Feynman integral $I_{a_1,\dots,a_n}$ is 
\begin{align}
I_{a_1 \ldots a_n}  &=  \frac{\mathcal{J} \; (-i)^L \, \pi^{(L-n)/2} \, \mathcal{E}^{(E-d+1)/2}}{\prod_{l=1}^L \Gamma \big( (d{+}1{-}E{-}l)/2 \big)} \int_{\mathsf{C}} \frac{ x_{n_P+1}^{-a_{n_P+1}} \cdots x_{n}^{-a_{n}}}{x_1^{a_1} \cdots x_{n_P}^{a_{n_P}}}  \, \mathcal{B}^{(d{-}E{-}L{-}1)/2} \, \bigwedge_{i=1}^{n}\id x_{i}
\,,
\label{eq:Baikov}
\end{align}
The integration variables are such that $x_1$ to $x_{n_P}$ are equivalent to the propagators of the original integral (this is the defining property of the Baikov representation), while the remaining $x_i$ correspond to irreducible scalar products, similarly to eq.~\eqref{eq:FI}. $L$ is the number of loops, $E$ the number of independent external momenta, and the two polynomials $\mathcal{B}$ (the Baikov polynomial) and $\mathcal{E}$ are Gram determinants
\begin{align}
\mathcal{E} = \det \! \big( G(p_1,\ldots,p_E) \big) \;, \quad\qquad \mathcal{B} = \det \! \big( G(k_1,\ldots,k_L,p_1,\ldots,p_E) \big)
\,.
\end{align}
Lastly $\mathcal{J}$ is the Jacobian for the variable change to the Baikov variables, and $\mathsf{C}$ is the integration region defined to be the region where $\mathcal{B}/\mathcal{E}$ is positive.

The Baikov representation is convenient due to the ease with which generalized cuts, as discussed in the next section, can be performed; they are simple residue operations with unit Jacobian. Furthermore, the IBP identities can also be reformulated in the Baikov representation. 
Specifically, eq.~\eqref{eq:IBP} becomes
\begin{align}
0 &= \int_{\mathsf{C}} \frac{\partial}{\partial x_j} \frac{ N(x) \mathcal{B}^{(d{-}E{-}L{-}1)/2} }{x_1^{a_1} \cdots x_{n_P}^{a_{n_P}}}  \, \bigwedge_{i=1}^{n}\id x_{i}
\,,
\end{align}
where $N(x)$ is an arbitrary polynomial in the $x_i$. 

\subsubsection*{IBPs in Lee--Pomeransky representation}

While less common, IBPs can be used to reduce Feynman integrals in \textit{Lee--Pomeransky representation}~\cite{Lee:2013hzt} (similar to the better known \textit{Feynman parameter representation}). This is another parametric representation
\begin{align}
    I_{a_{1},\dots,a_{n}} = \frac{(-1)^{a}\Gamma\left(\frac{d}{2}\right)}{\Gamma\left((L+1)\frac{d}{2}-a\right)\prod_{i}\Gamma(a_{i})}\int_{0}^{\infty}\mathcal{G}^{-\frac{d}{2}}\bigwedge_{i=1}^{n}x_{i}^{a_{i}}\frac{\dd x_{i}}{x_{i}}\,,
    \label{eq:leepomrep}
\end{align}
where $\mathcal{G}(x_{1},\dots,x_{n})$ is known as the Lee--Pomeransky polynomial and is given by
\begin{align}\label{eq:LP_Poly}
    \mathcal{G} = \mathcal{U}+\mathcal{F}\,.
\end{align}
Here, the polynomials $\mathcal{U}$ and $\mathcal{F}$ are the usual Symanzik polynomials. This representation naturally assumes that all $a_{i}>0$, which is not necessarily true for generic Feynman integrals. The representation used in ref.~\cite{Lu:2024dsb, Kreimer:2010gf} accounts for these cases, and can be summarised as
\begin{align}
    I_{a_{1},\dots,a_{n}} &= \frac{(-1)^{a}\Gamma\left(\frac{d}{2}\right)}{\Gamma\left((L+1)\frac{d}{2}-a\right)\prod_{a_{i}>0}\Gamma(a_{i})}J_{a_{1},\dots,a_{n}}\,,\\
    J_{a_{1},\dots,a_{n}} &:= \int_{0}^{\infty} \Bigg( \bigwedge_{a_{i}>0}x_{i}^{a_{i}}\frac{\dd x_{i}}{x_{i}} \Bigg) \Bigg(\bigwedge_{a_{i}\leq0}\delta(x_{i})dx_{i}(-\partial_{i})^{-a_{i}} \Bigg) \mathcal{G}^{-\frac{d}{2}} \nonumber \\
    &= \int_{0}^{\infty} \Bigg( \bigwedge_{a_{i}>0}x_{i}^{a_{i}}\frac{\dd x_{i}}{x_{i}} \Bigg) R_0 \, \mathcal{G}_{0}^{-\frac{d}{2}} \,.
\end{align}
where
\begin{align}
    \mathcal{G}_{0} := \mathcal{G}|_{x_i|_{a_i \leq 0} \rightarrow 0} \,, \qquad R_0 := \mathcal{G}^{d/2} \Bigg(\prod_{a_{i}\leq0} (-\partial_{i})^{-a_{i}} \Bigg) \mathcal{G}^{-\frac{d}{2}} \Big|_{x_i|_{a_i \leq 0} \rightarrow 0}
\end{align}

For the remainder of the paper, we will use the integrals $J_{a_{1},\dots,a_{n}}$ when working in Lee--Pomeransky representation.

With all $a_{i}>0$ \footnote{These IBPs are easily generalizable to the case with some $a_{i}\leq0$ by using $\mathcal{G}_{0}$ and removing these parameters.}, the core IBP identity in this representation~\cite{Lee:2014tja}\footnote{IBPs in Lee--Pomeransky representation, Feynman parameter representation, and related representations are not usually discussed due to the presence of these boundary terms, which are absent in momentum space and in Baikov representation. For some other past uses, see e.g. refs.~\cite{Sameshima:2019qpr, Artico:2023jrc, Lu:2024dsb}.} includes a boundary term
\begin{align}
    \int_{\Gamma}\dd \Big(\mathcal{G}^{-\frac{d}{2}}\phi_{k}\dd\widehat{x}_k \Big)+\int_{\partial\Gamma_{k}} \!\! \left(\mathcal{G}^{-\frac{d}{2}}\phi_{k}\right)_{\! x_{k}=0}\dd\widehat{x}_k=0\,,
\end{align}
where $\phi_{k}$ is some polynomial, $k\in[n]$ (with no $k$-sum implied), and where throughout this paper we will adopt the notation
\begin{align}
    \dd\widehat{x}_I = \bigwedge_{j \in [n]\setminus I} \dd x_j \,.
\label{eq:xhatdef}
\end{align}
This can be generalized by taking a sum of such identities
\begin{align}
    \int_{\Gamma}\dd\bigg(\sum_{k}\phi_{k}\mathcal{G}^{-\frac{d}{2}}\dd\widehat{x}_k\bigg)+\sum_{k}\int_{\partial\Gamma_{k}} \!\! \left(\phi_{k}\mathcal{G}^{-\frac{d}{2}}\right)_{\! x_{k}=0}\dd\widehat{x}_k&=0 \;\, \Leftrightarrow \label{eq:IBPinleepomrep}
\end{align}
\vspace{-5mm}
\begin{align}
    \int_{\Gamma}\sum_{k}\left(\partial_{k}\phi_{k}-\frac{d}{2 \mathcal{G}}\phi_{k}\partial_{k}\mathcal{G}\right)\mathcal{G}^{-\frac{d}{2}}\bigwedge_{i=1}^{n}\dd x_{i}+\sum_{k}\int_{\partial\Gamma_{k}} \!\! \left(\phi_{k}\mathcal{G}^{-\frac{d}{2}}\right)_{\! x_{k}=0}\dd\widehat{x}_k&=0\,.
\end{align}
The presence of $\mathcal{G}$ in the denominator results in $(d{+}2)$-dimensional integrals appearing in the IBP identities\footnote{IBPs in Baikov representation also introduce dimension-shifted integrals, and one must also solve syzygy equations to ensure these do not appear in the final identities \cite{Wu:2023upw}.}. It is often convenient to choose the polynomials $\{\phi_{k}\}$ such that we avoid this, and we can do that by enforcing the syzygy condition
\begin{align}\label{eq:syzygy_condition}
\phi_{0}\mathcal{G}+\sum_{k}\phi_{k}\partial_{k}\mathcal{G}=0\,,
\end{align}
for some additional polynomial $\phi_{0}$. The syzygies $\phi_{l}$ can be calculated using standard computer algebra tools such as \texttt{Singular}~\cite{DGPS}. The resulting IBP identity reads
\begin{align}\label{eq:LP_Syz_IBP}
    \int_{\Gamma}\bigg(\frac{d}{2}\phi_{0}+\sum_{k}\partial_{k}\phi_{k}\bigg)\mathcal{G}^{-\frac{d}{2}}\bigwedge_{i=1}^{n}\dd x_{i}=-\sum_{k}\int_{\partial\Gamma_{k}} \!\! \left(\phi_{k}\mathcal{G}^{-\frac{d}{2}}\right)_{\! x_{k}=0}\dd\widehat{x}_k\,.
\end{align}
We will make use of IBPs in this representation in \cref{sec:proof}.

\subsection{Sectors and spanning cuts}
\label{sec:sectors}

Here, we elaborate on the notion and significance of sectors.
Given an integral $I_{a_{1},\dots,a_{n}}$, its corresponding sector is defined as
\begin{align}\label{eq:sector}
    S = \{i\in[1,n_P]\,|\,a_{i}>0\}\,,
\end{align}
where $a_{n_{P}+1},\dots,a_{n}$ are excluded as they can never be positive. 
Sectors are an important organizational tool because, generically, IBP identities \eqref{eq:linear} only relate integrals within a given sector and its subsectors. 
More explicitly, let $S$ be the so-called generating sector corresponding to the exponents $\{a_1,\dots,a_{n_P}\}$ in eq.~\eqref{eq:IBP}.
Then, each integral $I_i$ in eq.~\eqref{eq:linear} either belongs to $S$ or to one of its subsectors $S^\prime \subset S$. This is because while taking the derivative in eq.~\eqref{eq:IBP} can raise the power of existing denominators, it can never introduce a denominator that was not there to begin with. 

A given integral family will have $2^{n_P}$ possible sectors, however many of these sectors are actually \textit{scaleless} and integrals therein evaluate to zero in dimensional regularization (dimreg). An example of a scaleless integral is the massless tadpole:
\begin{align}
    \int\frac{d^{d}k}{i\pi^{d/2}}\frac{1}{k^{2}}=0\,.
\end{align}
If a sector contains a scaleless integral, then that sector is also scaleless, and all integrals in this sector are zero in dimreg. There are many well understood algorithms for determining scaleless sectors \cite{Lee:2012cn,Maierhofer:2017gsa}.

Once the set of non-vanishing sectors is determined, we define the notion of \textit{spanning cuts} \cite{Larsen:2015ped}. In this work we will consider cuts in an analogous way to how they are considered in IBP reduction algorithms, as a contour deformation around the pole formed by the cut propagator, which is equivalent to a residue operation in Baikov representation~\cite{Baikov:1996iu, Frellesvig:2017aai}
\begin{align}\label{eq:residue}
    \int_{\mathsf{C}}\dd x\,\frac{\mathcal{B}^{\gamma}}{x^{n}} \;\; \rightarrow \;\; \frac{1}{2 \pi i} \oint\dd x\,\frac{\mathcal{B}^{\gamma}}{x^{n}} \,=\, \text{Res}_{x=0}\left(\frac{\mathcal{B}^{\gamma}}{x^{n}}\right)\,.
\end{align}

For the purpose of IBP reduction, it is sufficient to determine whether or not a given integral \textit{vanishes} on a cut. Consider taking a cut of a subset of propagators $C \subseteq \{1,\dots,n_P\}$. It is clear from the residue operation that the integral vanishes on this cut if at least one of the propagators in $C$ appears with a non-positive power (not as a denominator), namely\footnote{We note that this is a very \textit{ad hoc} way of applying cuts, but in practice this has been used extensively and successfully in IBP algorithms. However the outcome of this work would suggest that we need to rethink this notion of cuts.}
\begin{align}
    I_{a_{1},\dots,a_{n}} \Big|_{C} = 0
    \quad \text{if } a_{i}\leq 0 \text{ for any } i\in C \, .
\end{align}

Crucially, the IBP coefficients $c_{\vec a}$ in eq.~\eqref{eq:linear} are usually invariant under the cut operation. Thus,
\begin{align}
    0 = \sum_{\vec a} c_{\vec a} I_{\vec a}
    \quad \implies \quad
    0 = \sum_{\vec a} c_{\vec a} I_{\vec a}\Big|_{C} \, .
\end{align}
This is a consequence of IBPs being relations at the level of the integrand, so applying a residue operation will leave the relation invariant. The same cannot be said for symmetry relations, which rely on information from the contour. Working on a cut therefore significantly reduces the size of the IBP system. The reduction obtained on a given cut is related to the full reduction by
\begin{align}
    I\Big|_{C} = \sum_{i=1}^{\nu} c_{i} J_{i}\Big|_{C}
    \quad \implies \quad
    I = \sum_{i\in\sigma_{C}} c_{i} J_{i} + \dots \, ,
\end{align}
where $\sigma_{C}$ denotes the set of master integrals that do not vanish on the cut $C$. Equivalently, it labels the subset of coefficients $c_i$ that can be determined from this cut. The remaining coefficients can then be obtained by considering other cuts on which the corresponding master integrals contribute.

To determine the full set of coefficients $c_i$, one must ensure that the chosen cuts collectively cover all master integrals, or equivalently all non-zero sectors. This is achieved by constructing a \textit{spanning set of cuts} $\mathcal{Z}$. These are defined to be any set of cuts such that for every non-zero sector there exists a member of $\mathcal{Z}$ which is a subsector of it. Mathematically speaking, we have
\begin{align}
\forall S \in \mathcal{N}, \, \exists C \in \mathcal{Z} , \, C \subseteq S
\end{align}
where $\mathcal{N}$ is the set of non-zero sectors. 
It is clear that the choice of spanning cuts is not unique, however in practice it is optimal to choose the cuts from the set of non-zero sectors, $\mathcal{Z}\subseteq\mathcal{N}$, to both maximize the size of the cuts and minimize the number of cuts.

As an example, consider the one-loop massless box, defined by the propagators
\begin{align}
    P_{1} = k^{2}, \quad\;\; P_{2} = (k{-}p_{1})^{2}, \quad\;\; P_{3} = (k{-}p_{1}{-}p_{2})^{2}, \quad\;\; P_{4} = (k{-}p_1{-}p_2{-}p_3)^{2},
\end{align}
where 
\begin{align}
    p_{i}^{2}=0\,, \quad\;\; s = (p_{1}{+}p_{2})^2, \quad\;\; t = (p_{2}{+}p_{3})^2, \quad\;\; u = (p_{1}{+}p_{3})^{2}=-s-t\,.
\end{align}
In this case $n=n_{P}=4$, so there are no ISPs. Out of the $16$ sectors there are $7$ non-zero sectors:
\begin{align}
    \mathcal{N} = \big\{\{1,2,3,4\},\, \{1,2,3\},\, \{1,2,4\},\, \{1,3,4\},\, \{2,3,4\},\, \{1,3\},\, \{2,4\}\big\}\,.
\end{align}
From these the optimal set of spanning cuts is given by
\begin{align}
    \mathcal{Z} = \big\{\{1,3\},\, \{2,4\}\big\}\,.
\end{align}
There are $3$ master integrals in this family, which can be chosen as the box $I_{1,1,1,1}$ and the two bubbles $I_{1,0,1,0}$, $I_{0,1,0,1}$. The reduction of an integral (here shown as $I_{2,3,1,1}$) is then given by
\begin{align}
\adjustbox{raise=-7.2mm}{\includegraphics[width=16mm]{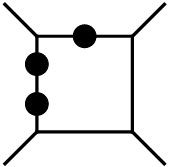}} \hspace{0mm} \, 
= \;\,
c_1 \times \!\!\!\adjustbox{raise=-7.2mm}{\includegraphics[width=16mm]{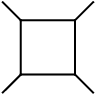}} \, + \;\, c_2 \times \adjustbox{raise=-1.8mm}{\includegraphics[width=17.5mm]{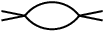}} \;\, + \;\, c_3 \times \, \adjustbox{raise=-7.3mm}{\includegraphics[height=16mm]{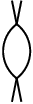}}
\end{align}
where the goal of the IBP reduction is to find the values of the $c_i$-coefficients. Yet the same coefficients can be extracted from simpler IBP reductions, i.e. reductions performed on a spanning set of cuts.
The two spanning cuts $C_{1}=\{1,3\}$ and $C_{2}=\{2,4\}$ give the reductions:
\begin{align}
\adjustbox{raise=-7.2mm}{\includegraphics[width=16mm]{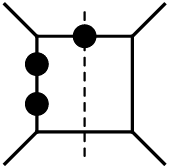}} \, &= \;\,
c_1 \times \!\!\!\adjustbox{raise=-7.2mm}{\includegraphics[width=16mm]{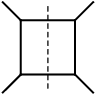}} \, + \;\, c_2 \times \adjustbox{raise=-3.8mm}{\includegraphics[width=17.5mm]{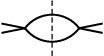}} \;\, + \qquad 0 \\[1mm]
\adjustbox{raise=-7.2mm}{\includegraphics[width=16mm]{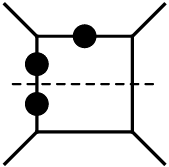}} \, &= \;\,
c_1 \times \!\!\!\adjustbox{raise=-7.2mm}{\includegraphics[width=16mm]{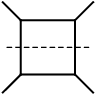}} \, + \,\hspace{13mm} 0 \hspace{12.5mm} + \;\, c_3 \times\adjustbox{raise=-7.3mm}{\includegraphics[height=16mm]{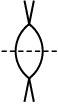}} \!\!
\end{align}
Thus all three IBP coefficients $c_{i}$ can be extracted by considering these two cuts, without ever performing the reduction of the complete system.

\subsection{Counting master integrals}
\label{sec:numberofmasters}

As discussed above, the space of Feynman integrals associated to a given graph forms a finite dimensional vector space. The size of this vector space and its basis elements---so-called master integrals---are important quantities in the study of Feynman integrals, and play an important role in the discussion of magic relations. The number of master integrals $\nu$ is a natural byproduct of performing an IBP reduction, while other approaches to integral reduction, such as those based on intersection theory~\cite{Mastrolia:2018uzb, Frellesvig:2019uqt}, need it as input. 
In general, one expects the space of Feynman integrals to be grouped according to sectors where each sector contains $\nu_S$ master integrals. Therefore, 
\begin{equation}\label{eq:nu}
    \nu = \sum_{S\,\in\,\text{sectors}} \nu_S\,.
\end{equation}
Aside from performing (potentially very expensive) IBP computations, there exist various methods from computational algebraic geometry to compute $\nu_S$, as well as $\nu$ directly \cite{Lee:2013hzt,Mastrolia:2018uzb,Fevola:2023fzn,Frellesvig:2019uqt,Caron-Huot:2021xqj}. 

For this work, we consider the computation of $\nu_S$ through the \textit{Lee--Pomeransky criterion}. The Lee--Pomeransky criterion has two variants, both discussed in \cite{Lee:2013hzt}.
However, as we will see, this counting becomes subtle in the presence of magic relations which mix sectors (mathematically, sectors are no longer a good filtration).

\subsubsection*{Counting in Lee--Pomeransky Representation}

Consider a Feynman integral in a sector $S$, which in Lee--Pomeransky representation can be written as
\begin{align}
I_{a_1,\ldots,a_n} &\propto \int_{\Gamma} \! \Phi_{S} \, \varphi_{a_1,\ldots,a_n}\,,
\label{eq:multivalued}
\end{align}
where $\varphi_{a_1,\ldots,a_n}$ is a rational $n$-form and $\Phi_{S}=\mathcal{G}_{S}^{-d/2}$ is a multivalued function shared by all integrals in the sector. Here, $\mathcal{G}_{S}$ represents the Lee--Pomeransky polynomial restricted to the sector $S$:
\begin{align}
    \mathcal{G}_{S} := \mathcal{G}\big|_{x_{i}\rightarrow0\,\forall\;i\notin S}\,.
\end{align}
In this representation, the \textit{restricted} Lee--Pomeransky criterion is given by:
\begin{align}
\nu_S \; = \; \text{\# of solutions to $\omega_{S} = 0$}\,,
\label{eq:restrictedleepom}
\end{align}
where 
$\omega_{S}$ is a one-form given by
\begin{align}\label{eq:crit_point_equations}
\omega_{S} := d \log(\Phi_{S}) = \sum_{i\in S}\omega_{S,i} \id x_i \qquad \text{with} \qquad \omega_{S,i} := \frac{\partial \log(\Phi_{S})}{\partial x_i}=-\frac{d}{2\,\mathcal{G}_{S}}\frac{\partial\, \mathcal{G}_{S}}{\partial x_i}
\,,
\end{align}
where the last equality holds for the Lee--Pomeransky representation specifically.

The restricted Lee--Pomeransky criterion can also be expressed in terms of the \textit{critical locus} $\text{Crit}(\omega_{S})$ as
\begin{align}
\crit(\omega_{S}) := \{\text{solutions to } \omega_{S} = 0\}\,, \qquad \nu_S \; = |\crit(\omega_{S})| := \# \text{ of elements in }\crit(\omega_{S})
\label{eq:restrictedleepomv2}
\end{align}

This criterion for counting master integrals is what is often quoted in the literature. However for certain particularly degenerate Feynman integrals, it can happen that $\crit(\omega_{S})$ is not comprised of isolated zero dimensional points. In such cases $\nu_S$ is not well defined through \cref{eq:restrictedleepom} and further analysis is required. This \textit{full} criterion is discussed in \cref{sec:criticalvarieties}.%
\footnote{%
One still has to be careful because the full criterion is supposed to be applied sector-by-sector and can over count if a super sector contains magic relations. 
}

\subsubsection*{Recasting in the Language of Polynomial Ideals}
The conditions $\omega_{S,i} = 0$ involve solving a system of rational equations. They can be recast as a \textit{polynomial} set of equations by adding a new equation, and variable $x_0$, as follows%
\footnote{Technically, $\mathcal{I}_{S}$ is an ideal in the ring $\mathbb{Q}(\text{kinematics})[x_0,\cdots,x_n]$.}
\begin{equation}\label{eq:euler_characteristic_ideal}
    \mathcal{I}_{S} := \left\langle \frac{\partial \mathcal{G}_{S}}{\partial x_1},\cdots, \frac{\partial \mathcal{G}_{S}}{\partial x_n},1-x_0 \mathcal{G}_{S}\right\rangle 
\end{equation}
Since $x_0 = \infty$ is not an allowed solution, the new constraint enforces that $\mathcal{G}_{S} \neq 0$. The advantage of working with \cref{eq:euler_characteristic_ideal} is that all entries are polynomials, allowing one to tackle the problem of finding critical points with the machinery of commutative algebra. In particular, it is possible to recast many of the concepts discussed above in the language of ideals. Indeed, the solution (vanishing) set to $\mathcal{I}_{S}=0$, denoted as $V(\mathcal{I}_{S})$, is equivalent to the critical variety\footnote{{Modulo the extra coordinate $x_0$ which can be safely ignored.}}
\begin{equation}
    V(\mathcal{I}_{S}) = \crit(\omega_{S})\,.
\end{equation}
Furthermore, if $\crit(\omega_{S})$ is comprised only of points, then the \textit{dimension} of the ideal $\dim(\mathcal{I}_{S}) = 0$, and the \textit{degree} of $\mathcal{I}_{S}$ is given by the number of such points (counting multiplicity):
\begin{equation}\label{eq:zero_dimensional_dim_relation}
    \deg(\mathcal{I}_{S}) = \nu_S, \,\, \text{if} \,\, \dim(\mathcal{I}_{S}) = 0\,.
\end{equation}
If instead $\crit(\omega_{S})$ contains a higher dimensional variety, then the $\textit{degree}$ of the ideal no longer corresponds to the number of solutions. This implies
\begin{equation}
    \deg(\mathcal{I}_{S}) \overset{?}\neq \nu_S, \,\, \text{if} \,\, \dim(\mathcal{I}_{S}) > 0\,.
\end{equation}

Importantly \cref{eq:zero_dimensional_dim_relation} implies that the number of master integrals can be computed without explicitly solving the system of equations given in \cref{eq:crit_point_equations}. Indeed the computation of ideal degrees and dimensions is a standard procedure implemented in many computer algebra systems such as \texttt{Singular}~\cite{DGPS} or \texttt{Macaulay2}~\cite{M2}, as well as a \textsc{Mathematica} implementation $\text{SP}\mathbb{Q}\text{R}$ \cite{Chestnov:2025svg}.

In the case of $\dim(\mathcal{I}_{S}) \neq 0$ then one can perform a \textit{minimal primary ideal decomposition} to separate the different components of $V(\mathcal{I}_{S})$:
\begin{equation}
    \mathcal{I}_{S} = \mathcal{I}_{S,1}\, \cap\,\cdots\,\cap\, \mathcal{I}_{S,m} \,.
    \label{eq:primary}
\end{equation}
Each primary component $\mathcal{I}_{S,i}$ has a respective vanishing set $V(\mathcal{I}_{S,i})$ of fixed dimension. This allows one to isolate and study the higher dimensional critical varieties more efficiently, as required for the full Lee--Pomeransky criterion discussed in \cref{sec:criticalvarieties}. Minimal primary ideal decompositions are standard procedures once again implemented in many computer programs such as \texttt{Singular} and \texttt{Macaulay2}.
\subsubsection*{Counting in Baikov Representation}
It is also possible to apply the technology discussed above to compute $\nu_S$ in Baikov representation, however the procedure is initially different. To compute the number of master integrals in a sector $\nu_{S}$ the procedure is to first take the \textit{maximal cut} of the Baikov polynomial with respect to the sector $S$:
\begin{equation}
    \mathcal{B}_{S} := \mathcal{B}\big|_{x_{i}=0\,\forall\;i\in S}
\end{equation}
The procedure is then identical to above with $\mathcal{G}_{S}\rightarrow\mathcal{B}_{S}$, only this time the variables are $x_{i}$ with $i \notin S$.

\subsubsection*{Regulated Counting}

It is also possible to directly find the total number of master integrals $\nu$, without having to iterate and sum over each sector's $\nu_S$ contribution (c.f., \eqref{eq:nu}). This method is often referred to as ``regulated counting" \cite{Frellesvig:2019kgj,Fontana:2023amt,Brunello:2023rpq}. 
For this we need to choose a top sector $S$ of the integral, such that all sectors that we are interested in are subsectors thereof. In Lee--Pomeransky, one then considers the twist
\begin{align}
    \Phi_{S,\mathrm{reg}} = \mathcal{G}_{S}^{-d/2}\prod_{i\in S}x_{i}^{\rho_{i}}\,.
\end{align}
For generic $d$ and $\rho_i$, the critical loci of the regulated twisted $\crit(\omega_\mathrm{S,reg}:=\dd\log\Phi_{S,\mathrm{reg}})$ is guaranteed to be a set of isolated points. 
The number of master integrals is then the degree of the resulting ideal $\mathcal{I}_\mathrm{S,reg}$ which is equal to the number of critical points: 
\begin{align}
    \nu = \sum_{\sigma\subseteq S}\nu_{\sigma} = \deg(\mathcal{I}_{S,\mathrm{reg}}) = |\crit(\omega_{S,\mathrm{reg}})|\,.
    \label{eq:nuregcount}
\end{align}
Note that the notion of a sector does not exist for Feynman integrals regulated by generic $\rho_i$. Therefore the given top sector $S$ and all of its subsectors are counted at once, and magic relations arising from these sectors appear as normal IBP identities.

In Baikov representation the strategy is almost identical; one builds the regulated twist
\begin{align}
    \Phi_{S,\mathrm{reg}} = \mathcal{B}^{(d-E-L-1)/2} \prod_{i\in S} x_{i}^{\rho_i}\,, 
\end{align}
and computes the degree of the associated ideal/counts the number of critical points.

\subsection{Symmetry relations}
\label{sec:symmetry}

Symmetry relations are a different type of relation between Feynman integrals. They are not relations at the level of the cohomology (i.e., the integrand), which is why they can not be obtained from IBPs, and are only valid given the additional information contained in the integration contour.

Symmetry relations come in two main types, symmetries \textbf{within} a sector, and symmetries \textbf{between} sectors. An example of each is
\begin{align}
& \; \text{Symmetry within a sector:} & \hspace{-1.5cm} \adjustbox{raise=-2.8mm}{\includegraphics[width=0.12\textwidth]{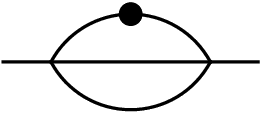}} \,&=\, \adjustbox{raise=-2.8mm}{\includegraphics[width=0.12\textwidth]{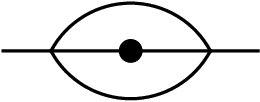}} \qquad \label{eq:symmetry1} \\
& \text{Symmetry between sectors:}    & \hspace{-1.5cm} \adjustbox{raise=-7.3mm}{\includegraphics[width=0.11\textwidth]{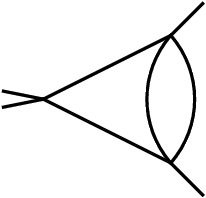}} \;&=\; \adjustbox{raise=-7.3mm}{\includegraphics[width=0.11\textwidth]{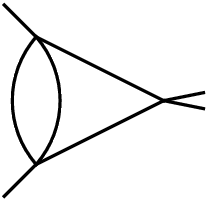}} \qquad
\end{align}
where the dots in eq.~\eqref{eq:symmetry1} indicate doubled propagators.
While not all symmetry relations fall neatly into these two categories (see \cite{Becchetti:2025rrz, Coro:2025vgn} for recent discoveries of more intricate symmetry relations), it is only the two above types that will concern us in this work.

If applied naively, symmetry relations may cause the spanning cuts method to fail. This is because symmetry relations only hold true at the integral level, and thus are altered by a change of contour (such as a cut). Symmetries must thus be applied only between sectors that survive on a given cut, or their use postponed to after the cut analysis.

After the application of symmetry relations, the number of master integrals often drops quite drastically. In this work we mainly concern ourselves with the counting of master integrals without the inclusion of these symmetries, see ref.~\cite{Duhr:2026elp} for recent efforts into counting the number of master integrals in the presence of these.
\section{Magic relations and critical varieties}
\label{sec:magicrelations}

There are cases where the methods discussed in the previous section, namely IBP-identities, critical point counting, and spanning cuts do not work as expected. In this section we introduce two such cases: \textit{magic relations} for IBP identities, and the emergence of \textit{higher dimensional critical varieties} when counting master integrals. Finally we conclude with the central observation of this work: namely that magic relations and higher dimensional critical varieties imply each other.

\subsection{Magic relations}
\label{sec:magic}

In \cref{sec:sectors}, we argued that IBP identities relate integrals that either belong to the generating sector or its subsectors. 
There are cases, however, where all integrals in the final IBP relation belong to subsectors, meaning that all integrals in the generating sector have coefficient $0$. Whenever an IBP relation of that type cannot be obtained from combining IBPs generated by subsectors, that IBP relation is known as a \textit{\textbf{magic relation}}~\cite{Maierhofer:2018gpa}. We thus define a magic relation to have the following properties:

\begin{myboxtitle}{\hypertarget{box:magicDef}{Magic relations} }
\begin{enumerate}
    \item The generating sector cannot appear in the identity, only its subsectors can appear.
    \item It cannot be obtained by combining IBPs generated in the subsectors themselves.
\end{enumerate}
\end{myboxtitle}
\newcommand{\magicDef}[1]{\hyperlink{box:magicDef}{#1}}

While the term ``magic relations'' was coined in ref.~\cite{Maierhofer:2018gpa}, they have been studied earlier, e.g. in refs.~\cite{Smirnov:2013dia, Lee:2010ik}.

\begin{figure}
\centering
\vspace{-2mm}
\includegraphics[width=4.5cm]{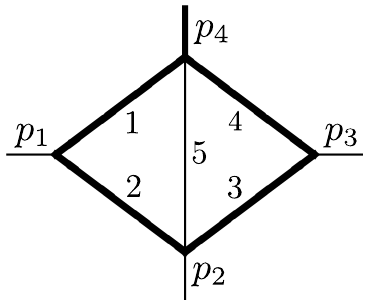}
\vspace{-2mm}
\caption{Generating sector for the Higgs magic relation. All external momenta are outgoing.}
\label{fig:higgsgen}
\end{figure}

A concrete example is best to illustrate how magic relations live up to their name.
Consider the Feynman integral family whose top sector is depicted on \cref{fig:higgsgen}. It may be parametrized by
\begin{align}
P_1 &= k_1^2 - m_t^2\,, & P_2 &= (k_1{-}p_1)^2 - m_t^2\,, & P_3 &= (k_2{-}p_1{-}p_2)^2 - m_t^2\,, \nonumber \\
P_4 &= (k_2{+}p_4)^2 - m_t^2\,, & P_5 &= (k_1{-}k_2)^2\,; & P_6 &= (k_1{-}p_1{-}p_2)^2 - m_t^2\,, \nonumber \\
P_7 &= k_2^2 - m_t^2\,, & P_8 &= (k_1{+}p_4)^2 - m_t^2\,, & P_9 &= (k_2{-}p_1)^2 - m_t^2\,,
\label{eq:higgspara}
\end{align}
where $P_1$-$P_5$ are genuine propagators while $P_6$-$P_9$ are ISPs. The kinematics is given by
\begin{align}
p_1^2 = p_2^2 = p_3^2 &= 0 \,,\quad\; p_4^2 = m_H^2\,, \nonumber \\
(p_1{+}p_2)^2 = s \,,\quad\; (p_2{+}p_3)^2 &= t \,,\quad\; (p_1{+}p_3)^2 = m_H^2 {-} s {-} t\,.
\label{eq:higgskinematics}
\end{align}
This family has been studied in ref.~\cite{Bonciani:2016qxi}\footnote{In the study in ref.~\cite{Bonciani:2016qxi} this relation was overlooked, giving rise to an over counting of one master integral in what in that paper is called \textit{Family A}.} in the context of Higgs-plus-jet production at NLO.

Within this family, there is the IBP identity
\begin{align}
s \, I_{11101;0000} + (m_H^2{-}s) I_{10111;0000} &= t \, I_{01111;0000} + (m_H^2{-}t) I_{11011;0000}
\,,
\label{eq:magichiggs}
\end{align}
or graphically
\begin{align} s \times \! \adjustbox{raise=-5mm}{\includegraphics[height=12.5mm]{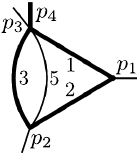}} \, + \, (m_H^2{-}s) \times \! \adjustbox{raise=-5mm}{\includegraphics[height=12.5mm]{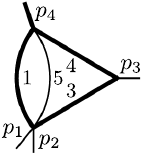}} \; = \; t \times \! \adjustbox{raise=-5mm}{\includegraphics[height=12.5mm]{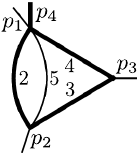}} \, + \, (m_H^2{-}t) \times \! \adjustbox{raise=-5mm}{\includegraphics[height=12.5mm]{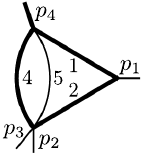}}
\,.
\label{eq:magichiggs2}
\end{align}
At a first glance, the above identity seems unlikely to be true since the four terms each depend on different sectors. Furthermore --- despite all containing integrals of ``triangle-bubble''-type --- each term depends on different subsets of the external kinematic variables $\{s, t, m_H^2\}$. 
This makes it appear somewhat ``magical'' and it is indeed an example of a magic relation. Concretely this identity can be derived from IBPs in the five-propagator sector $I_{11111;0000}$ with the IBP relation:
\begin{align}
0 &= \int \frac{d^d k_1 d^d k_2}{- \pi^d} \left( \frac{\partial}{\partial k_1^{\mu}} \frac{p_1^{\mu} (P_8-P_6)}{P_1 P_2 P_3 P_4 P_5} + \frac{\partial}{\partial k_2^{\mu}} \frac{p_3^{\mu} (P_7-P_9)}{P_1 P_2 P_3 P_4 P_5} \right)
\end{align}
followed by IBPs generated in the various sub-sectors.

\subsection{Magic relations and cuts}
\label{sec:magicandcuts}

Magic relations are of special interest since much of the common wisdom~\cite{Primo:2016ebd, Primo:2017ipr, Bosma:2017ens, Frellesvig:2017aai} about Feynman integrals breaks down in their presence. Particularly, IBPs are no longer invariant under the cut operation described in~\cref{eq:residue} in the presence of magic relations.
To see this, recall eqs.~\eqref{eq:magichiggs} and \eqref{eq:magichiggs2}. 
Cutting the propagators indexed $\{1,2,3,5\}$ yields
\begin{align}
s \times \adjustbox{raise=-5.9mm}{\includegraphics[height=13.5mm]{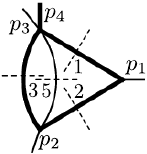}} \; \stackrel{?}{=} \; 0
\,,
\label{eq:magiconcut}
\end{align}
since only the first of the four terms of eq.~\eqref{eq:magichiggs} has support on this cut. Eq.~\eqref{eq:magiconcut} implies that this integral must vanish. 
However, the $\{1,2,3,5\}$ cut of $I_{11101;0000}$ is non-zero, which is explicitly computed in Appendix \ref{app:cut_calc}. Therefore \cref{eq:magiconcut} is not valid. Thus the cutting procedure described in \cref{sec:sectors} may fail in the presence of magic relations and can thus restrict the choice of spanning cuts. Naively, the optimal choice for the spanning cuts would be
\begin{align}
    \mathcal{Z} = \big\{\{1,3\},\{1,4\},\{2,3\},\{2,4\}\big\}\,.
\end{align}
However, in practice this set of spanning cuts fails due to the magic relation. The only safe choice of cut would be one where all sectors involved in the magic relation survive, which would mean only cutting the 5th propagator $\{5\}$. With this in mind, the only safe choice for the spanning cuts ends up being no cuts at all:
\begin{align}
    \mathcal{Z} = \big\{\{\}\big\}
\end{align}

Cuts and magic relations interfering can also have an influence on the counting of master integrals in a sector related to other sectors through a magic relation, when performed using the Lee--Pomeransky criterion of \cref{sec:numberofmasters} on the maximal cut of the sector. A couple of examples of this are discussed in ref.~\cite{Frellesvig:2019kgj}.

\subsubsection*{Relative twisted cohomology}

Another place where magic relations cause problems is intersection theory~\cite{Mastrolia:2018uzb, Frellesvig:2019uqt} in the context of \textit{relative twisted cohomology}~\cite{Caron-Huot:2021xqj, Caron-Huot:2021iev, Brunello:2023rpq, Crisanti:2024onv, Lu:2024dsb, De:2023xue, De:2024zic, Glew:2025ypb}. 
In this approach (from the point of view of the Baikov representation which is how it is discussed in ref.~\cite{Brunello:2023rpq}), dual forms to the differential form defining the given master integral, have to contain cutting operators corresponding to the integral sector. These operators, known as \textit{delta-forms}, work such that the intersection pairing localizes on all singularities in the differential form unregulated by the twist. 
What should happen in practice that if the $n$-form $\varphi$ has $n_u$ unregulated poles, the $n$-variate intersection number of $\varphi$ with a corresponding delta-form will reduce to the $(n{-}n_u)$-variate intersection number of $\text{Res}_{n_u \!}(\varphi)$ with the coefficient of the delta-form. But when the integral of $\varphi$ is related, through magic relations, to other integrals, such a residue operator cannot be straightforwardly defined, and the relative approach to intersection theory appears to lose its utility\footnote{In \cite{Caron-Huot:2021xqj}, the appearance of a magic relation at one-loop was identified with the existence of non-trivial non-middle dimensional cohomology groups.}.

\subsection{Magic relations and symmetries}
\label{sec:magicandsymmetry}

As discussed in \cref{sec:magic} all magic relations will, by definition, relate integrals in subsectors of the generating sector. They may, however, \textit{appear} quite differently after the application of symmetry relations (see \cref{sec:symmetry}). Since in practical calculations in phenomenology symmetries and IBPs usually are applied together, this fact may have played a role in obscuring the common origin of magic relations. 

We may sort magic relations into three different \textit{symmetry types} depending on their appearance after the use of symmetries:

\textbf{Type A}: Magic relations that (even after the use of symmetry relations) relate integrals in different sectors.

\textbf{Type B}: Magic relations that, after the use of symmetry relations, relate integrals in one sector (and potentially subsectors thereof) and appear as if they were an extra IBP relation in that sector.

\textbf{Type C}: Magic relations that are directly implied by symmetry relations, and therefore become trivial after the application of symmetries.

The Higgs example from \cref{sec:magic} is an example of symmetry type A, and in \cref{sec:examples} we will show a number of additional examples of magic relations, and we will encounter examples of each of the three kinds, to be summarized at the end of that section.

\subsection{Critical varieties and the full Lee--Pomeransky criterion}
\label{sec:criticalvarieties}
In \cref{sec:numberofmasters}, we introduced the restricted Lee--Pomeransky criterion, where the number of master integrals in a given sector $S$ was given by the number of solutions to $\dd\log\Phi_{S}=0$. It may happen, however, that solving this equation does not only result in isolated critical points, but also in a critical line or other forms of higher dimensional critical varieties. In that case $\log(\abs{\Phi_{S}})$ is (by definition) no longer a Morse function, and the proof of the restricted Lee--Pomeransky criterion (as shown, for instance, in appendix~\ref{app:morse}) no longer holds.

This case was also considered in ref. \cite{Lee:2013hzt}. 
The prescription given there, for computing the number master integrals, is to count critical points \textit{on} each critical surface. 
In other words, we have what we call the \textbf{full Lee--Pomeransky criterion}
\begin{align}
    \nu_S = \sum_{i}\, \text{number of critical points on $\mathcal{C}_{S,i}$}\,,
    \label{eq:fullleepom}
\end{align}
where the $\mathcal{C}_{S,i}$ denote the critical varieties of the components in the minimal primary ideal decomposition of the critical variety $\crit(\omega_{S}) = \bigcup_i \mathcal{C}_{S,i}$.
Whenever the $\mathcal{C}_{S,i}$ contain only isolated points, this criterion simply reduces to the restricted criterion of eq.~\eqref{eq:restrictedleepom}. The argument for counting the critical points of the $\mathcal{C}_i$ in ref.~\cite{Lee:2013hzt}%
\footnote{In ref.~\cite{Lee:2013hzt} the full Lee--Pomeransky criterion is justified with a reference to ref.~\cite{ArnoldSingularities}.} parallels that of Morse--Bott theory (see appendix~\ref{app:morsebott}).

Suppose the Gr{\"o}bner basis for the ideal responsible for the higher dimensional critical variety has at most one non-linear generator $\tilde{\mathcal{P}}$. What is done in practice~\cite{Lee:2013hzt} is to construct the new ideal
\begin{align}
\tilde{\mathcal{I}} = \left\langle \partial_{x_1} \tilde{\mathcal{P}} ,\; \ldots ,\; \partial_{x_n} \tilde{\mathcal{P}} ,\; 1 - x_0 \tilde{\mathcal{P}} \right\rangle
\end{align}
and count the number of critical points contained in $\tilde{\mathcal{I}}$.

Despite being more refined when compared to the restricted Lee--Pomeransky criterion, there are still many subtleties in both cases where one has to be careful. For example, counting master integrals sector by sector will sometimes overcount the true number of master integrals due to possible magic relations appearing from super sectors. Furthermore, it is also possible for this method to undercount due to the possibility of critical points moving to infinity in the limit of vanishing analytic regulators \cite{Boels:2015yna,Bitoun:2017nre}. 

Regulated counting (see \cref{sec:numberofmasters}) has the advantage that neither of these problems occur because the notion of sectors is lost. For this reason it provides a more robust framework for computing $\nu$. Nevertheless, the introduction of the analytic regulators significantly complicates the structure of the critical point equation systems, rendering the computation of their respective solution count significantly more computationally challenging.

\subsubsection*{Example of a higher dimensional critical variety}
Let us revisit the double-triangle discussed for the Higgs example in \cref{sec:magic}, shown in \cref{fig:higgsgen}, and compute the number of master integrals in the top sector $S=\{1,2,3,4,5\}$. To find $\nu_S$ we must consider $\Phi_{S} = \mathcal{G}_{S}(x_1,\ldots,x_5)^{-d/2}$, allowing us to compute the ideal
\begin{align}
\mathcal{I}_{S} = \big\langle \partial_{x_1} {\mathcal{G}_{S}},\, \ldots,\, \partial_{x_5}{\mathcal{G}_{S}},\; 1 - x_0 {\mathcal{G}_{S}} \big\rangle\,.
\end{align}
By performing a minimal primary ideal decomposition, $\mathcal{I}_{S}$ factors into an intersection of three primary ideals as
\begin{align}
\mathcal{I}_{S} = \mathcal{I}_{S,1} \cap \mathcal{I}_{S,2} \cap \mathcal{I}_{S,3}
\end{align}
These ideals are given by
\begin{align}
\mathcal{I}_{S,1} &= \Big\langle F_1^2 x_5 {+} 2 F_2 u \,,\;\; F_1 x_4 {-} 2 s \,,\;\;
F_1 x_1 {-} 2 t \,,\;\;
F_1 x_2 {+} 2 (m_H^2{-}s) \,,\;\;
F_1 x_3 {+} 2 (m_H^2{-}t) \,,\;\;
4 s t u^2 x_0 {-} F_1^3 \Big\rangle\,, \nonumber \\
\mathcal{I}_{S,2} &= \Big\langle 3 s t x_5+2 u \,,\;\;
x_4 (m_H^2{-}t)+s x_3 \,,\;\;
3 m_t^2 s u x_2 - (m_H^2{-}s) (3 m_t^2 u x_4 - s) \,, \nonumber \\
&\qquad\qquad\qquad st - 3 m_t^2 u (s x_1 {+} t x_4) \,,\;\;
27 m_t^2 s t-4 u x_0 \,,\;\;
9 m_t^2 t u x_4^2 - 3 s t x_4 + 2 s \Big\rangle \,, \\
\mathcal{I}_{S,3} &= \Big\langle x_5 \,,\;\;
1 + 3 m_t^2 (x_3 + x_4) \,,\;\;
1 + 3 m_t^2 (x_1 + x_2) \,,\;\;
x_0-27 m_t^4 \,,\;\; \mathcal{Q} \, \Big\rangle \,, \nonumber
\end{align}
where we have defined
\begin{align}
u := m_H^2 {-} s {-} t \,, \qquad\;\;\; F_1 := 4 m_t^2 u + st \,, \qquad\;\;\; F_2 :=  2 m_t^2 u + st\,,
\end{align}
and
\begin{align}
\mathcal{Q} := 3 m_t^2 (m_H^2{-}s) x_4 + 3 m_t^2 \big( 3 m_t^2 u \, x_4 - s \big) x_2 - 2 m_t^2 - s\,.
\label{eq:Ptildehiggs}
\end{align}
The zero-locus for $\mathcal{I}_{S,1}$ is given by the point
\begin{align}
    x_1= \frac{2 t}{F_1} \,,\quad\; x_2= -\frac{2 \big(m_H^2 {-} s \big)}{F_1} \,,\quad\; x_3= -\frac{2 \big(m_H^2 {-} t \big)}{F_1} \,,\quad\; x_4= \frac{2 s}{F_1} \,,\quad\; x_5= -\frac{2 F_2 u}{F_1^2}\,,
\end{align}
where here and in the following the solution for $x_0$ is omitted. The zero locus for $\mathcal{I}_{S,2}$ is given by the two conjugate critical points
\begin{align}
    x_1= \frac{s t \pm \sqrt{\Delta }}{6 m_t^2 s u}\,,\;\;\,
    x_2= \frac{\big(s{-}m_H^2\big) \big(s t \pm \sqrt{\Delta }\big)}{6 m_t^2 s t u}\,,\;\;\,
    x_3= \frac{\big(t{-}m_H^2\big) \big(s t \mp \sqrt{\Delta} \big)}{6 m_t^2 s t u}\,,\;\;\,
    x_4= \frac{s t\mp\sqrt{\Delta }}{6 m_t^2 t u}\,,\;\;\,
    x_5= \frac{- 2 u}{3 s t}\,,
\end{align}
with $\Delta := s t \left(s t-8 m_t^2 u\right)$.
Lastly the zero locus for $\mathcal{I}_{S,3}$ is given by
\begin{align}
    x_2= \frac{-1}{3 m_t^2}-x_1 \,,\quad\;
    x_3= \frac{m_t^2 \big( 3 (m_H^2{-}t) x_1 - 2 \big)-t}{3 m_t^2 ( t - 3 m_t^2 x_1 u)} \,,\quad\;
    x_4= \frac{2-3 s x_1}{3 ( t - 3 m_t^2 x_1 u )} \,,\quad\;
    x_5= 0 \,,
    \label{eq:higgshyperbola}
\end{align}
where $x_1$ is an unfixed degree of freedom. This is an example of a one-dimensional critical variety. 
\subsubsection*{Example of the full Lee--Pomeransky criterion}

If we want to count master integrals in a sector containing a critical variety, such as the one given by eq.~\eqref{eq:higgshyperbola}, we can follow the prescription given by the full Lee--Pomeransky criterion of eq.~\eqref{eq:fullleepom} and look at the non-linear generator of $\mathcal{I}_{S,3}$. That is the polynomial given by \cref{eq:Ptildehiggs}, which we thus identify as $\tilde{\mathcal{P}} := \mathcal{Q}$.
Solving simultaneously the equations $\partial_{x_2} \tilde{\mathcal{P}} = 0$, $\partial_{x_4} \tilde{\mathcal{P}} = 0$, and $1 - x_0 \tilde{\mathcal{P}}=0$ results in one critical point:
\begin{align}
x_2 = \frac{s-m_H^2}{3 m_t^2 u} \,,\quad x_4 = \frac{s}{3 m_t^2 u}\,.
\end{align}
Combining everything, one solution is obtained from $\mathcal{I}_{S,1}$, two from $\mathcal{I}_{S,2}$, and one from $\mathcal{I}_{S,3}$ (after the second iteration), therefore $\nu_S=4$. This result is in agreement with public IBP codes such as FIRE~\cite{Smirnov:2025prc}, as well as~\cite{Bonciani:2016qxi} where these integrals were computed.

\subsection{Critical varieties and magic relations are equivalent}
\label{sec:equivalence}
In the previous sections, the same Feynman integral was used both to showcase a magic relation, and as an example of a critical variety. This connection is not a coincidence and it provides the first example of the main result of this paper: 
\begin{myboxtitle}[label={box:equivalent}]{\hypertarget{box:keyObs}{Key observation}}
    \centering
    A sector $S$ generates a magic relation
    $\;\Leftrightarrow\;$
    $S$ has a higher dimensional critical variety.
\end{myboxtitle}
\newcommand{\keyObs}[1]{\hyperlink{box:keyObs}{#1}}
More specifically, we observe that if IBPs in a certain sector generate magic relations between its subsectors, then a computation of $\nu_S$ for that sector will result in a higher dimensional critical variety, and vice versa. 
We have examples of this in both directions, i.e. both examples where the magic relation was known before identifying the critical variety, as well as examples where the critical variety was known before finding the magic relation. A detailed list of examples is given in \cref{sec:examples}.

\section{Connecting higher-dimensional critical varieties to magic relations}
\label{sec:proof}

In this section we will demonstrate how the dimensionality of the critical variety generated by $\mathcal{I}$ is intrinsically connected to the existence of magic relations. In particular, we argue the following statement:
\begin{align}
    \boxed{
    \text{A sector $S$ has only zero dimensional critical varieties} 
    \; {\Rightarrow} \;
    \text{$S$ cannot generate a magic relation}
    }
\label{eq:equivalent2}
\end{align}
In order for our argument to hold, we assume that the Lee--Pomeransky polynomial $\mathcal{G}$, as defined by~\cref{eq:LP_Poly}, satisfies
\begin{equation}\label{eq:gcd_condition}
    \partial_i\, \text{gcd}(\partial_{i}\mathcal{G},\partial_{j}\mathcal{G})=\partial_j\, \text{gcd}(\partial_{i}\mathcal{G},\partial_{j}\mathcal{G})=0\,,
\end{equation}
for all $i,j=1,\dots,n$, $i\neq j$. In words, the pairwise greatest common divisor of $\partial_{i}\mathcal{G}$ and $\partial_{j}\mathcal{G}$ must be a constant in the $i,j$ variables.

This is a technical assumption needed to rule out a degenerate class of multiplier functions. We are not aware of a Feynman integral with $\dim(\mathcal{I})=0$ that does not satisfy this condition. Whether the dimensionality of $\mathcal{I}$ in addition to the structure of $\mathcal{G}$ is sufficient to always enforce this condition, is left to further investigation.

For notational convenience we will consider the sector $S=[n] := \{1,\ldots,n\}$ with Feynman parameters $x_{i}$, $i=1,\dots,n$, and drop the $S$ subscript for the remainder of this section. 

\subsection{The Koszul complex and its cohomology}

In order to understand the connection between the dimensionality of the critical variety and the existence of magic relations, we turn to the language of commutative algebra, using the \textit{Koszul complex}. 

The Koszul complex characterizes how the map $\omega\wedge$ acts on the space of differential $p$-forms on $X=\mathbb{C}^n \setminus V(\mathcal{G})$, denoted by $\Omega^p(X)$.
More explicitly, the Koszul complex is the following sequence of spaces connected by the map $\omega\wedge$ 
\begin{align}
    0 \rightarrow 
    \Omega^0(X) \xrightarrow{\omega\wedge} 
    \Omega^1(X) \xrightarrow{\omega\wedge} 
    \cdots \xrightarrow{\omega\wedge} 
    \Omega^{n-2}(X) \xrightarrow{\omega\wedge} 
    \Omega^{n-1}(X) \xrightarrow{\omega\wedge} 
    \Omega^n(X) \xrightarrow{\omega\wedge}  
    0
    \,. 
\end{align}
Here, $\omega\wedge: \Omega^{p-1}(X) \to \Omega^{p}(X)$ maps a $(p-1)$-form to a $p$-form.
Because $\omega\wedge$ is nilpotent (i.e., squares to zero: $\omega\wedge \omega \wedge \bullet = 0$) and the image, $\mathrm{im}(\omega\wedge: \Omega^{p-1}(X) \to \Omega^{p}(X))$, is contained in the kernel, $\mathrm{ker}(\omega\wedge: \Omega^{p}(X) \to \Omega^{p+1}(X))$, it is often called a ``derivative''. We refer to it as the \textit{generalized derivative} to not confuse it with the usual derivative $\dd$.

A $p$-form $\varphi^{(p)}\in\Omega^{p}(X)$ is \textit{exact} under this generalized derivative if there exists a $(p-1)$-form $\varphi^{(p-1)}\in\Omega^{p-1}(X)$ such that $\varphi^{(p)} = \omega\wedge\varphi^{(p-1)}$, and it is \textit{closed} if $\omega\wedge\varphi^{(p)}=0$. The space of closed forms modulo exact forms under a generalized derivative is called the \textit{cohomology}, and measures the failure of $\Omega^{p}(X)$ to be exact. The Koszul cohomology is thus given by
\begin{align}
    H^{p}_\mathrm{K} (X;\omega)&:= \frac{\{\text{Closed forms on }\Omega^{p}(X)\}}{\{\text{Exact forms on }\Omega^{p}(X)\}} = \frac{
        \text{ker}(\omega\wedge: \Omega^{p}(X) \rightarrow \Omega^{p+1}(X)
        )
    }{
        \text{im}(\omega\wedge: \Omega^{p-1}(X) 
        \rightarrow \Omega^{p}(X))
    } 
    \,.
\end{align} 
This is very sensitive to the critical locus of $\Phi=\mathcal{G}^{-\frac{d}{2}}$ (recall that $\omega = \dd\log \Phi$) because this is exactly where $\omega$ vanishes: $\crit(\omega) = \{x^* \in X : \omega(x^*) = 0\}$. This means that on the critical locus, $\omega\wedge$ maps all forms to zero.

The Koszul cohomology can also be thought of as a first step in computing the twisted cohomology which uses the twisted derivative $\nabla_\omega = \dd + \omega\wedge$.
\footnote{The twisted cohomology is the cohomology of the Koszul cohomology with respect to the usual exterior derivative $\dd$. More formally, the Koszul cohomology appears on the first page of a spectral sequence that converges to the twisted cohomology \cite{Mizera:2019vvs, Matsubara-Heo:2023hmf}.}
The top-degree cohomology group $H_{K}^{n}$ has a direct interpretation in terms of the master integral counting. Away from $\mathcal{G}=0$, we expect
\begin{align}\label{eq:HKn}
    H^{n}_\mathrm{K}= \frac{
        \{f\; \mathrm{d}^n x : f\in \Omega^0(X)\}
    }{
        \{ \bigl(\sum_{i=1}^n \omega_i \phi_i\bigr) \mathrm{d}^n x : \phi_i \in \Omega^0(X)\}
    }
    =
    \frac{\mathbb{C}[x_0,x_1,\cdots,x_n]}{\left\langle 
        1-x_0 \mathcal{G}
        ,\, \partial_1 \mathcal{G}
        ,\,\cdots,\,
        \partial_n \mathcal{G}
    \right\rangle}
    = \frac{\mathbb{C}[\vec{x}]}{\mathcal{I}}
    \,.
\end{align}
When $\dim(\mathcal{I})=0$ the dimension of this quotient ring equals $\dim H_K^n = \deg(\mathcal{I}) = \nu_S$, the number of master integrals. 
This gives an alternative perspective of the restricted Lee--Pomeransky criterion: the number of master integrals is the dimension of $H^n_\mathrm{K}$.%
\footnote{%
    This is reasonable because when $\mathrm{dim}(\mathcal{I})=0$, the twisted cohomology and Koszul cohomology are ``almost the same''.
    More precisely, when $\mathrm{dim}(\mathcal{I})=0$, the twisted cohomology degenerates to the Koszul cohomology
    in the large dimension limit ($d\to \infty$) \cite{Mizera:2019vvs}. 
    Furthermore, despite not being isomorphic as vector spaces, $\mathrm{dim}H^n(X,\nabla_\omega) = \mathrm{dim}H^n_{\mathrm{K}}(X,\omega)$ and a basis of $H^n_{\mathrm{K}}(X,\omega)$ can be uplifted to a basis of $H^n(X,\nabla_\omega)$ \cite{Matsubara-Heo:2023hmf}.
}

\subsection{The Koszul cohomology as syzygies}

We now consider the degree-($n{-}1$) cohomology group $H^{n-1}_\mathrm{K}$, which has an equally natural interpretation. 
An arbitrary $(n{-}1)$-form%
\footnote{We choose convenient signs for our forms to make their interpretation clearer.}
\begin{align}\label{eq:n-1form}
    \psi = \sum_{i=1}^n (-1)^{i-1}\phi_i\,\mathrm{d}\hat{x}_i\,,
\end{align}
where $\hat{x}_i$ is defined as in eq.~\eqref{eq:xhatdef}, is closed under $\omega\wedge$ if and only if $\omega\wedge\psi = 0$
\begin{align}\label{eq:closedsyzygy}
    \omega\wedge\psi = \left(\sum_{i=1}^n \omega_i \phi_i\right)\mathrm{d}^n x = 0 \qquad \Longleftrightarrow \qquad \sum_{i=1}^n \frac{\partial_i\mathcal{G}}{\mathcal{G}}\phi_i = 0\,.
\end{align}
Away from $\mathcal{G}=0$ this is exactly the syzygy condition \eqref{eq:syzygy_condition} with $\phi_0=0$
\begin{equation}\label{eq:syzygy_no_phi0}
    \sum_{i=1}^{n}\phi_i\partial_i\mathcal{G} = 0\,.
\end{equation}
Similarly, $\psi$ is exact if it lies in the image of $\omega\wedge: \Omega^{n-2}\to\Omega^{n-1}$.
\begin{align}\label{eq:KoszulExactnessInCoDeg2}
    \psi &= \omega \wedge
    \overset{\in \Omega^{n-2}(X)}{\overbrace{
        \sum_{\substack{i,j\in[n] \\ i<j}} 
        (-1)^{i+j-1} \phi_{ij}\, 
        \mathrm{d}\hat{x}_{ij}
    }}
    \iff 
    \phi_m = -\sum_{l<m} \omega_l \phi_{lm} 
        + \sum_{l>m} \omega_l \phi_{ml}
    \,.
\end{align}
These are forms made by taking combinations of the $\omega_{i}$, and they contain the set of \textit{trivial syzygies}. Trivial syzygies can be defined as solutions whose vanishing set contains $V(\mathcal{I})$ fully, and they are discussed in more detail in appendix~\ref{app:trivialargument}. The cohomology group $H_{K}^{n-1}$ can be interpreted as 
\begin{align}
    H_{K}^{n-1} = \frac{\{\text{syzygies of~\cref{eq:syzygy_no_phi0}}\}}{\{\text{trivial syzygies}\}}\,.
\end{align}
We refer to the elements of this group as \textit{non-trivial} syzygies. We will go on to conjecture that these are exactly the generators of magic relations.

\subsection{Zero dimensional critical varieties imply trivial syzygies}
\label{sec:zerodimtrivial}

When the critical locus consists only of isolated points ($\dim(\mathcal{I})=0$), a theorem in commutative algebra (see e.g.\ \cite[Theorem 17.6]{Eisenbud_1995}) implies that the Koszul complex is exact in all degrees except $p=n$, meaning
\begin{align}
    \dim(\mathcal{I}) = 0 \qquad \Longrightarrow \qquad H^p_\mathrm{K} = 0 \quad \text{for all } p < n\,.
\end{align}
In particular $H^{n-1}_\mathrm{K}=0$, which by the identification above means there are no non-trivial syzygies.

Conversely, when $\dim(\mathcal{I})>0$, the Koszul complex fails to be exact in lower degrees, and $H^{p<n}_\mathrm{K}$ becomes non-trivial. 
The elements of $H^{n-1}_\mathrm{K}$ are the non-trivial syzygies that we conjecture generate magic relations. We further expect that
\begin{align}
    \dim H_{K}^{n-p}>0, \quad p=0,\dots,\dim(\mathcal{I})\,.
\end{align}
We leave the interpretation of the elements of $H_{K}^{n-p}$ for $p>1$ for future work.

This statement can be intuitively understood as follows: Consider a syzygy solution of \cref{eq:syzygy_no_phi0}:
\begin{align}\label{eq:syz_inverse}
    \sum_{k=1}^{n}\phi_{k}\partial_{k}\mathcal{G}=0 \;\; \implies \;\; \partial_n \mathcal{G} 
    =-\frac{1}{\phi_n}\,\sum_{k=1}^{n-1}\,\phi_{k}\partial_{k}\mathcal{G}\,. 
\end{align}
This equation can be naively substituted into $\mathcal{I}$ to reduce the number of independent equations from $n+1$ to $n$. Since there are $n+1$ variables to be solved for, \cref{eq:syz_inverse} implies that $\dim(\mathcal{I})\neq0$. The only way for this \textit{not} to occur is for \cref{eq:syz_inverse} to be invalid on the support of $V(\mathcal{I})$:
\begin{equation}\label{eq:trivial_syzygy_def}
    \phi_n(x^*) = 0\,,\quad x^* \in V(\mathcal{I}) \;\; \implies \;\; \phi(x^*) = 0 \quad \text{if} \quad \dim(\mathcal{I}) = 0\,,
\end{equation}
thereby showing that all syzygy solutions to \cref{eq:syzygy_no_phi0} must be trivial if $\dim(\mathcal{I})=0$. From this argument we would also expect that
\begin{align}\label{eq:numberofmagic}
    \dim H_{K}^{n-1} = \dim(\mathcal{I}) = \{\#\text{ of non-trivial syzygies of \cref{eq:syzygy_no_phi0}}\}\,,
\end{align}
namely that the number of non-trivial syzygies, or the dimension of the cohomology group, is equivalent to the dimension of the critical variety.

\subsection{Magic relations are non-trivial syzygies}

We have argued that a higher dimensional critical variety results in the existence of non-trivial syzygy solutions to \cref{eq:syzygy_no_phi0}. Following from~\cref{eq:LP_Syz_IBP}, these generate IBPs of the form:
\begin{align}\label{eq:LP_Syz_IBP_2}
    \int_{\Gamma}\left(\sum_{k=1}^{n}\partial_{k}\phi_{k}\right)\mathcal{G}^{-\frac{d}{2}}\bigwedge_{i=1}^{n}\dd x_{i}=-\sum_{k=1}^{n}\int_{\partial\Gamma_{k}} \!\! \left(\phi_{k}\mathcal{G}^{-\frac{d}{2}}\right)_{\! x_{k}=0}\dd\widehat{x}_k\,.
\end{align}
Recall from \cref{sec:magic} that two important criteria were presented to define a \hyperlink{box:magicDef}{magic relation}. The \textit{first} criterion was that the generating sector drops out of the relation. For the top sector contribution to vanish, the additional condition
\begin{align}
    \sum_{k=1}^{n}\partial_{k}\phi_{k}=0
    \label{eq:div_free_cond}\,.
\end{align}
must thus be enforced. We will refer to these as \textit{divergence-free} syzygies, because the divergence of the vector $\vec\phi$ vanishes: $\nabla\cdot\vec\phi=0$.

The \textit{second} criterion of a \hyperlink{box:magicDef}{magic relation}, that the relation cannot be generated from the subsectors themselves, is more difficult to show. In appendix~\ref{app:trivialargument}, we provide an argument showing that trivial syzygies, under the assumption of the gcd condition of \cref{eq:gcd_condition}, do not produce magic relations. The implication of this statement is that magic relations can only be generated from non-trivial syzygies, and therefore
\begin{align}
    \{\text{magic relations}\}\subseteq \{\text{non-trivial syzygies}\} = H_{K}^{n-1}\,.
\end{align}
To recap, we have shown that if the critical variety is zero dimensional, $\dim(\mathcal{I})=0$, then the Koszul cohomology group $H_{K}^{n-1}$ is also zero dimensional. This implies that there are no non-trivial syzygy solutions of \cref{eq:syzygy_no_phi0}, and the only syzygy solutions that could possibly generate a magic relation must be trivial ones. Finally, appendix~\ref{app:trivialargument} argued that, with the gcd condition in~\cref{eq:gcd_condition}, trivial syzygies do not produce magic relations, therefore implying that no magic relations can be generated when we have a zero dimensional critical variety, which is equivalent to the statement in \cref{eq:equivalent2}.

We provide proof of concept \textsc{Mathematica} implementations for detecting and finding magic relations based on these principles in the ancillary file \texttt{Magic-Test.m}. These include a \texttt{MagicQ} function that detects whether a magic relation is expected to exist, and a \texttt{FindMagicRelations} function that attempts to find the corresponding magic relation. These rely on the $\text{SP}\mathbb{Q}\text{R}$, \textsc{Calico}, and \textsc{FiniteFlow} packages \cite{Chestnov:2025svg,Bertolini:2025zud,Peraro:2019svx}, and assume that the conjectured \hyperlink{box:keyObs}{Key observation} is correct and that the gcd condition in~\cref{eq:gcd_condition} is satisfied.

There are still many unanswered questions here: Is the gcd condition in~\cref{eq:gcd_condition} always implied by $\dim(\mathcal{I})=0$ and the general structure of $\mathcal{G}$? Are all non-trivial syzygies equivalent to magic relations? From our studies we have not found any independent non-trivial syzygies that are not divergence-free, do these exist and if so do they generate a different type of magic relation? These are all questions for future work.

\subsection{The Higgs example revisited}
\label{sec:Higgspart3}

As an example of the ideas mentioned throughout this section in action, let us revisit the Higgs family of sections~\ref{sec:magic} and \ref{sec:criticalvarieties}, shown in \cref{fig:higgsgen}. As discussed in \cref{sec:criticalvarieties}, the Lee--Pomeransky polynomial $\mathcal{G}$ has a one-dimensional critical variety in the top sector given by
\begin{align}
    \vec x = \left(x_{1} \,,\;\; -\frac{1}{3m_{t}^{2}}-x_{1} \,,\;\; \frac{t+m_{t}^{2}(2+3(t-m_{H}^{2})x_{1})}{3m_{t}^{2}(3m_{t}^{2}ux_{1}-t)} \,,\;\; \frac{3sx_{1}-2}{3(3m_{t}^{2}ux_{1}-t)} \,,\;\; 0\right)\,.
\end{align}
By solving the syzygy condition \cref{eq:syzygy_condition}, we find there is exactly one independent divergence free non-trivial syzygy, as expected:
\begin{align}
    \vec\phi = \Big(0\,,\;\; (s{-}m_{H}^{2})x_{1}-tx_{2} \,,\;\; (m_{H}^{2}{-}s)x_{1}+tx_{2} \,,\;\; (t{-}m_{H}^{2})x_{4}-sx_{3} \,,\;\; sx_{3}+(m_{H}^{2}{-}t)x_{4} \,,\;\; 0 \Big)\,.
\end{align}
This syzygy generates the following IBP identity in LP representation:
\begin{align}\label{eq:magic_IBP_higgs}
    -tJ_{0,2,1,1,1}+(m_{H}^{2}{-}s)J_{2,0,1,1,1}-(m_{H}^{2}{-}t)J_{1,1,0,2,1}+sJ_{1,1,2,0,1} = 0\,.
\end{align}
After further IBP reduction in all subsectors, we find that this identity is equivalent to \cref{eq:magichiggs}.
The above results are easily reproduced using the codes in the ancillary file:
\\
\begin{minipage}[c]{\linewidth}
\begin{lstlisting}
higgsGpol = (x1 + x2) (x3 + x4) (1 + mt^2 (x1 + x2 + x3 + x4)) + (mt^2 x1^2 + x2 +
     x3 + x4 - t x2 x4 + mt^2 (x2 + x3 + x4)^2 + 
    x1 (1 - s x3 - mh^2 x4 + 2 mt^2 (x2 + x3 + x4))) x5;
    
MagicQ[higgsGpol,{x1,x2,x3,x4,x5}]
(*True*)

FindMagicRelations[higgsGpol,{x1,x2,x3,x4,x5}]
(*{-tj[0, 2, 1, 1, 1]+(-mh^2+t)j[1, 1, 0, 2, 1]+sj[1, 1, 2, 0, 1]+(mh^2-s)j[2, 0, 1, 1, 1]}*)
\end{lstlisting}
\end{minipage}

\section{Examples}
\label{sec:examples}

We further motivate the observation of \cref{sec:criticalvarieties} by providing an extensive list of examples of Feynman integral sectors that generate magic relations and contain higher dimensional critical varieties. For each example we will show the magic relation and the critical variety, and we will count critical points \textit{on} the critical variety to test the full Lee--Pomeransky criterion. At the end of the section our findings are summarized. All Lee--Pomeransky representations discussed in this section are relevant for the top sector of the considered diagram, so we drop the $S$ subscript for convenience.

\subsection{Tadpole examples}
\label{sec:examplestadpoles}

We first consider two simple tadpole examples, which we call $T_{i}$. The first of these is the simplest example of a magic relation and we will go through it in some detail. The second is constructed from the first to have a critical variety of dimension larger than one.

\subsubsection*{Example $T_{1}$} 

The first example we consider is the bubble with two identical internal masses and, crucially, no external mass i.e. $p^2 = 0$. It is given as
\begin{align}
I_{a_1 a_2} &= \int \frac{d^d k}{i \pi^{d/2}} \frac{1}{P_1^{a_1} P_2^{a_2}}
\end{align}
where the two propagators are
\begin{align}
P_1 = k^2 - m^2 \,, \qquad\quad P_2 = (k-p)^2 - m^2 \,.
\end{align}
In this integral sector we can generate the IBP
\begin{align}
0 &= \int \frac{d^d k}{i \pi^{d/2}} \frac{\partial}{\partial k^{\mu}} \frac{p^{\mu}}{P_1 \, P_2}
\end{align}
which, after the use of additional IBPs in the individual subsectors, corresponds to the relation
\begin{align}
I_{10} &= I_{01}
\label{eq:tadpolemagic}
\end{align}
which is a magic relation. In addition to \cref{eq:tadpolemagic} being a magic relation, it is also a symmetry relation. Thus \cref{eq:tadpolemagic} is an example of what in \cref{sec:magicandsymmetry} is called type C.

For comparison we can consider the case where we do not have $p^2=0$. In that case eq.~\eqref{eq:tadpolemagic} is still true, as it still is a valid symmetry relation, but it will no longer be a magic relation since no IBP in the $p^2 \neq 0$ bubble sector can reproduce it.

To investigate the higher dimensional critical variety, both Lee--Pomeransky and Baikov representations are considered for this example.

For the Lee--Pomeransky representation we have
\begin{align}
    \mathcal{G} := \mathcal{U} + \mathcal{F} = (x_1 + x_2) \big( 1 + m^2 ( x_1 + x_2 ) \big)
\end{align}
By solving for $\dd \log( \mathcal{G}^{-d/2} ) = 0$ there exists one line-like solution given by
\begin{align}\label{eq:bubble_line}
x_1 = \frac{-1}{2 m^2} - x_2\,.
\end{align}
This can be confirmed using the provided ancillary file \texttt{Magic-Test.m} with
\\
\begin{minipage}[c]{\linewidth}
\begin{lstlisting}
MagicQ[(x1+x2)(1+m^2(x1+x2)),{x1,x2}]
(*True*)
FindMagicRelations[(x1+x2)(1+m^2(x1+x2)),{x1,x2}]
(*{-j[0, 1] + j[1, 0]}*)
\end{lstlisting}
\end{minipage}
\\[.5em]
The critical line in~\cref{eq:bubble_line} does not have any critical points and therefore by the full Lee--Pomeransky criterion (c.f.,~\cref{sec:criticalvarieties}), there are no master integrals in the generating sector.  
This corresponds to the known fact that a bubble with no external mass is reducible to tadpoles.

The Baikov parametrization for this integral is a little less straight forward, due to the fact (discussed in appendix B1 of ref.~\cite{Frellesvig:2024ymq}) that Baikov parameterizing an integral with massless two-point kinematics requires embedding it in a larger sector with less trivial kinematics. For that purpose we introduce the third propagator, playing the role of an ISP, 
\begin{align}
P_3 = (k+q)^2 - m^2 \qquad \text{with} \qquad q^2 = 0 \qquad \text{and} \qquad p \cdot q = c/2
\end{align}
With this the Baikov polynomial is
\begin{align}
\mathcal{B} &= (x_1 - x_2)(x_1 - x_3) + c (x_1 + m^2)
\end{align}
Cutting $x_1$ and $x_2$, the two propagators of the generating sector, leaves the Baikov polynomial a constant $\mathcal{B}|_{\text{cut}} = m^2 c$, meaning that the equation $d \log( \mathcal{B} ) = 0$ becomes
\begin{align}
0 &= 0
\label{eq:zeroiszero}
\end{align}
``Solving'' this equation for the remaining Baikov variable $x_3$ gives a line in $x_3$-space, which is how the critical variety appears in Baikov representation. This can be be once again confirmed with
\\
\begin{minipage}[c]{\linewidth}
\begin{lstlisting}
maxCut = ReplaceAll[{x1->0, x2->0}];
MagicQ[(x1-x2)(x1-x3)+c(x1+m^2) // maxCut, {x3}]
(*True*)
\end{lstlisting}
\end{minipage}
\\[.5em]
Presumably the triviality of eq.~\eqref{eq:zeroiszero} is the reason this critical line has not been discussed much in the past.

We should, however, note that this example has been considered previously, both as a magic relation~\cite{Wang:2024hsm, Lu:2024dsb} and as a critical variety~\cite{Page:2025gso}, though, to our knowledge, the connection has not been drawn.

\subsubsection*{Example $T_{2}$} 
Let us now consider an extension of example $T_1$, by ``squaring'' it. The generating sector will have the four propagators
\begin{align}
P_1 &= k_1^2 - m^2 \,, & P_2 &= (k_1-p)^2 - m^2 \,, & P_3 &= (k_2-p)^2 - m^2 \,, & P_4 &= k_2^2 - m^2 \,, \quad
\end{align}
and the three ISPs
\begin{align}
\quad P_5 &= (k_2+q)^2 - m^2 \,, & P_6 &= (k_1+q)^2 - m^2 \,, & P_7 &= (k_1-k_2)^2 \,. \qquad
\end{align}
The kinematics is
\begin{align}
p^2 = 0 \,,\qquad q^2 = 0 \,,\qquad (p+q)^2 = c \,.
\end{align}
On the maximal cut the Baikov polynomial becomes $c^2 x_7 (x_7 - 4 m^2)$, meaning that solving the critical point equations fixes $x_7 = 2 m^2$ but leaves $x_5$ and $x_6$ unfixed giving us a two-dimensional critical variety. From our argument in~\cref{sec:proof}, it would be natural to suggest that this implies that there are two non-trivial, divergence-free syzygies, and therefore two magic relations. This is indeed the case, and the two magic relations are of the form
\begin{align}
    I_{0111}-I_{1011} = 0 \,, \qquad\quad I_{1101}-I_{1110} = 0\,.
\end{align}
Both of these can be further IBP reduced from IBPs in their own sector to recover the ``square'' of the relation from $T_1$.

\subsection{FIRE examples}
\label{sec:examplesfire}

\begin{figure}
\centering
\adjustbox{raise=0mm}{\includegraphics[width=0.19\textwidth]{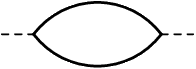}} \hspace{3mm}
\adjustbox{raise=1.0mm}{\includegraphics[width=0.26\textwidth]{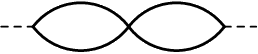}} \hspace{3mm}
\adjustbox{raise=-1.0mm}{\includegraphics[width=0.19\textwidth]{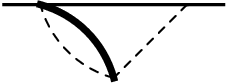}} \hspace{3mm}
\adjustbox{raise=-0.7mm}{\includegraphics[width=0.19\textwidth]{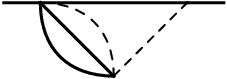}}
\caption{Generating sectors for the examples from sections \ref{sec:examplestadpoles} and \ref{sec:examplesfire}. From left to right are $T_1$, $T_2$, $S_1$, and $S_2$. Normal lines have mass $m$, dotted lines are massless, and the bold line has mass $M$.}
\label{fig:smsectors}
\end{figure}

Next we will discuss two examples of magic relations that were among the first to be discovered, and which were discussed in ref.~\cite{Smirnov:2013dia} in connection with the publication of FIRE4. We refer to these examples as $S_{i}$, and their generating sectors are depicted to the right on \cref{fig:smsectors}.

\subsubsection*{Example $S_1$} 
The first example from ref.~\cite{Smirnov:2013dia} has a generating sector described by the propagators
\begin{align}
P_1 &= k_1^2 \,, & P_2 &= (k_1-k_2)^2 - M^2 \,, & P_3 &= (k_2-p)^2 - m^2 \,, \nonumber \\
P_4 &= k_2^2 \,; & P_5 &= (k_1-p)^2 \,,
\end{align}
where $P_5$ is an ISP. The kinematics is such that $p^2 = m^2$. The corresponding magic relation is given as
\begin{align}
(8{-}3d) I_{11100} \, + \, 4 m^2 I_{11200} \, + \, 2 M^2 I_{12100} \, + \, (d{-}2) I_{11010} \,=\, 0
\end{align}
or graphically
\begin{align}
(8{-}3d) \, \adjustbox{raise=-3mm}{\includegraphics[width=0.10\textwidth]{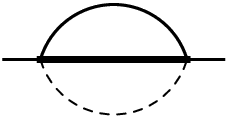}} \, + \,
4 m^2 \, \adjustbox{raise=-3mm}{\includegraphics[width=0.10\textwidth]{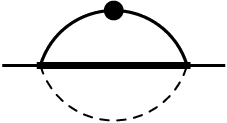}} \, + \,
2 M^2 \, \adjustbox{raise=-3mm}{\includegraphics[width=0.10\textwidth]{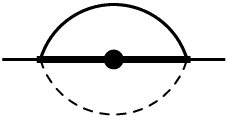}} \, + \,
(2{-}d) \, \adjustbox{raise=-3mm}{\includegraphics[width=0.067\textwidth]{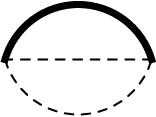}} \; &= \; 0
\label{eq:sm2}
\end{align}
The Baikov of the generating sector yields a constant on the cut, making $d \log(\mathcal{B}) = 0$ become $0=0$ so its solution (in the one remaining variable) is one dimensional, just as in the tadpole example.

\subsubsection*{Example $S_2$} 
The second of the examples from ref.~\cite{Smirnov:2013dia}, which first was discussed in ref.~\cite{Lee:2010ik}, is generated by the last diagram in \cref{fig:smsectors}. The problem may be parametrized as
\begin{align}
P_1 &= k_1^2 \,, & P_2 &= k_2^2 - m^2 \,, & P_3 &= (k_1{+}k_2{+}k_3)^2 - m^2 \,, \nonumber \\
P_4 &= (k_3{+}p)^2 - m^2 \,, & P_5 &= k_3^2 \,; & P_6 &= (k_1{+}p)^2 \,, \\
P_7 &= (k_2{+}p)^2 - m^2 \,, & P_8 &= (k_1{+}k_3)^2 \,, & P_9 &= (k_2{+}k_3)^2 \,, \nonumber
\end{align}
where $P_1-P_5$ are actual propagators and $P_6-P_9$ are ISPs. The kinematics is such that $p^2 = m^2$.

The corresponding magic relation is given as
\begin{align}
2(5{-}2d) \, I_{11110;0000} \, + \,
8m^2 \, I_{12110;0000} \, + \,
(d{-}2) \, I_{11101;0000} \; &= \; 0
\label{eq:sm1}
\end{align}
or graphically
\begin{align}
2(5{-}2d) \, \adjustbox{raise=-3mm}{\includegraphics[width=0.10\textwidth]{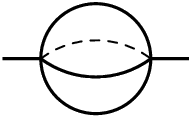}} \, + \,
8m^2 \, \adjustbox{raise=-3mm}{\includegraphics[width=0.10\textwidth]{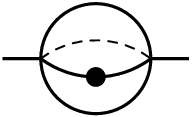}} \, + \,
(d{-}2) \, \adjustbox{raise=-3mm}{\includegraphics[width=0.06\textwidth]{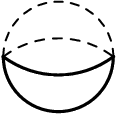}}  \; &= \; 0
\label{eq:sm1gr}
\end{align}
(in refs.~\cite{Smirnov:2013dia, Lee:2010ik} the relation is written differently due to a different choice of master integral basis).

Performing a Baikov parametrization, doing the maximal cut of the generating sector, and solving the equation system $d \log( \mathcal{B} ) = 0$, gives one solution which is the line parametrized as
\begin{align}
x_7 = \frac{3 m^2 - x_6}{2} \,,\quad x_8 = \frac{8 m^2}{3} \,, \quad x_9 = \frac{-m^2}{3} \,.
\end{align}
That line itself has no critical points, so we conclude that there are no master integrals in the generating sector, in agreement with FIRE.

\subsection{$g{-}2$ QED examples}
\label{sec:examplesqed}

Next we will discuss seven examples where the existence of a critical variety was known, but where we discovered the associated magic relation. Those examples are taken from ref.~\cite{Lee:2013hzt} and are all subsectors of a certain four-loop integral relevant for $g{-}2$ in QED. Out of the hundreds of subsectors, only these seven were found to contain such a critical variety. We refer to these examples as $L_{i}$.

\subsubsection*{Example $L_1$} 
The first of these is in ref.~\cite{Lee:2013hzt} known as $\#246$. We may use a parametrization with the propagators
\begin{align}
P_1 &= k_4^2\,, & P_2 &= k_1^2 - m^2\,, & P_3 &= k_2^2 - m^2\,, \nonumber \\
P_4 &= (k_3 + k_4 + p)^2 - m^2\,, & P_5 &= (k_1 + k_3)^2 - m^2\,, & P_6 &= (k_2 + k_3)^2 - m^2\,;
\end{align}
and the ISPs
\begin{align}
P_7 &= k_3^2\,, & P_8 &= (k_1 + k_2)^2\,, & P_9 &= (k_1 + k_4)^2 \,, & P_{10} &= (k_2 + k_4)^2\,, \nonumber \\ P_{11} &= (k_3 + k_4)^2\,, & P_{12} &= (k_1 + p)^2 \,, & P_{13} &= (k_2 + p)^2\,, & P_{14} &= (k_4 + p)^2\,.
\end{align}
The kinematics is such that $p^2 = m^2$.
The corresponding magic relation is given as
\begin{align}
\adjustbox{raise=-4mm}{\includegraphics[width=0.10\textwidth]{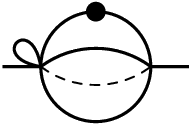}} \, - \, 
\adjustbox{raise=-4mm}{\includegraphics[width=0.10\textwidth]{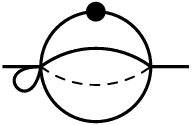}} \, - \, 
\adjustbox{raise=-4mm}{\includegraphics[width=0.10\textwidth]{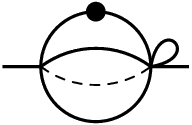}} \, + \, 
\adjustbox{raise=-4mm}{\includegraphics[width=0.10\textwidth]{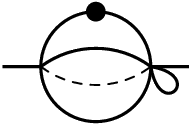}} \;\;\;\;\;\;\, \nonumber \\
- \,
\adjustbox{raise=-4mm}{\includegraphics[width=0.10\textwidth]{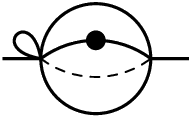}} \, + \, 
\adjustbox{raise=-4mm}{\includegraphics[width=0.10\textwidth]{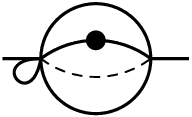}} \, + \, 
\adjustbox{raise=-4mm}{\includegraphics[width=0.10\textwidth]{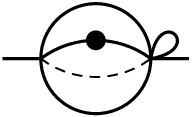}} \, - \, 
\adjustbox{raise=-4mm}{\includegraphics[width=0.10\textwidth]{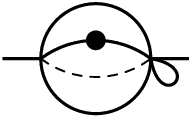}}
 = 0
\end{align}
which we can see becomes trivial (i.e. $0=0$) after the application of (either one of three kinds of) symmetry relations within or between the sectors.

We may then Baikov parametrize, do the maximal cut and solve $d \log( \mathcal{B}) = 0$, and the result is the two conjugate point-like solutions
\begin{align}
x_7 &= 2 m^2\,, & x_8 &= 3 m^2\,, & x_9 &= 2 m^2\,, & x_{10} &= 2 m^2\,, \nonumber \\
x_{11} &= 0\,, & x_{12} &= \tfrac{1}{2} \big( 5 \mp \sqrt{5} \big) m^2 \,, & x_{13} &= \tfrac{1}{2} \big( 5 \mp \sqrt{5} \big) m^2\,, & x_{14} &= \big( 2 \mp \sqrt{5} \big) m^2 \,,
\end{align}
along with the one-dimensional solution(s)
\begin{align}
x_7 &= 0\,, \quad\;\; x_8 = 2 m^2\,, & x_{10} &= \tfrac{1}{3} \left( - m^2 \mp \sqrt{79 m^4 {-} 6 m^2 x_9 {-} 9 x_9^2} \right), &
x_{11} &= \tfrac{8}{3} m^2\,, \nonumber \\
x_{12} &= \tfrac{1}{2} (5 m^2 - x_9)\,, & x_{13} &= \tfrac{1}{6} \left( 16 m^2 \pm \sqrt{79 m^4 {-} 6 m^2 x_9 {-} 9 x_9^2} \right), & x_{14} &= -\tfrac{1}{3} m^2\,,
\end{align}
parametrized in terms of $x_9$. We easily realize that these two solutions belong to the same critical variety. The nontrivial generator of the primary ideal responsible for that one-dimensional solution is
\begin{align}
\tilde{\mathcal{P}} = 710 m^4 - 6 m^2 x_{10} - 9 x_{10}^2 + 54 m^2 x_9 + 81 x_9^2
\end{align}
which itself has the one critical point
\begin{align}
x_9 = x_{10} = \frac{- m^2}{3}
\end{align}
so we conclude that the generating sector of $L_1$ has three master integrals in total (before applying symmetry relations), in agreement with FIRE~\cite{Smirnov:2025prc}.

\begin{figure}
\centering
\adjustbox{raise=0.3mm}{\includegraphics[width=0.125\textwidth]{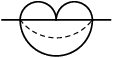}} \hspace{1mm}
\adjustbox{raise=0.3mm}{\includegraphics[width=0.125\textwidth]{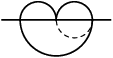}} \hspace{1mm}
\adjustbox{raise=0mm}{\includegraphics[width=0.125\textwidth]{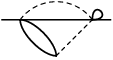}} \hspace{1mm}
\adjustbox{raise=0.5mm}{\includegraphics[width=0.125\textwidth]{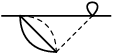}} \hspace{2mm}
\adjustbox{raise=-1.6mm}{\includegraphics[width=0.078\textwidth]{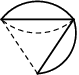}} \hspace{2mm}
\adjustbox{raise=0.3mm}{\includegraphics[width=0.125\textwidth]{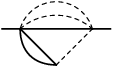}} \hspace{1mm}
\adjustbox{raise=0.3mm}{\includegraphics[width=0.125\textwidth]{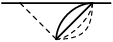}}
\vspace{-1mm}
\caption{Generating sectors for the seven magic relations $L_1$--$L_7$ whose critical varieties were discussed in ref.~\cite{Lee:2013hzt}. Solid lines have mass $m$ and dotted lines are massless.
\label{fig:sevensectors}}
\end{figure}

\subsubsection*{Example $L_2$} 
The second of these is in ref.~\cite{Lee:2013hzt} known as $\#350$. We will use the parametrization with the propagators
\begin{align}
P_1 &= k_2^2 - m^2\,, & P_2 &= (k_1 - k_2)^2 - m^2\,, & P_3 &= k_3^2 - m^2\,, \nonumber \\
P_4 &= k_4^2 - m^2\,, & P_5 &= (k_1 - k_3 - k_4)^2\,, & P_6 &= (k_1 - p)^2 - m^2\,;
\end{align}
and the ISPs
\begin{align}
P_7 &= k_1^2\,, & P_8 &= (k_1 - k_3)^2 - m^2\,, & P_9 &= (k_1 - k_4)^2 - m^2\,, & P_{10} &= (k_2 - k_3)^2\,, \nonumber \\
P_{11} &= (k_2 - k_4)^2\,, & P_{12} &= (k_2 - p)^2\,, & P_{13} &= (k_3 - p)^2\,, & P_{14} &= (k_4 - p)^2\,.
\end{align}
The kinematics is such that $p^2 = m^2$.
The corresponding magic relation is
\begin{align}
(5-2d) \Bigg( 
\adjustbox{raise=-4mm}{\includegraphics[width=0.10\textwidth]{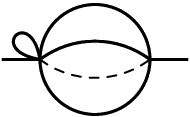}} - 
\adjustbox{raise=-4mm}{\includegraphics[width=0.10\textwidth]{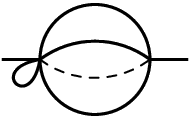}} + 
\adjustbox{raise=-4mm}{\includegraphics[width=0.06\textwidth]{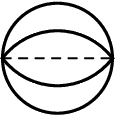}} \Bigg) + 4 m^2 \Bigg(
\adjustbox{raise=-4mm}{\includegraphics[width=0.10\textwidth]{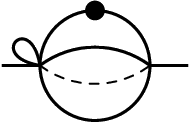}} -
\adjustbox{raise=-4mm}{\includegraphics[width=0.10\textwidth]{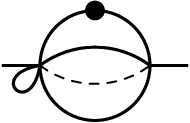}} +
\adjustbox{raise=-4mm}{\includegraphics[width=0.06\textwidth]{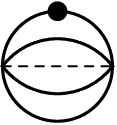}}
\Bigg) = 0
\end{align}
which after imposing symmetry relations between the sectors reduces to
\begin{align}
\adjustbox{raise=-4mm}{\includegraphics[width=0.06\textwidth]{figures/lp350_s6.eps}} \; &= \; \frac{2d-5}{4 m^2} \,
\adjustbox{raise=-4mm}{\includegraphics[width=0.06\textwidth]{figures/lp350_s5.eps}}
\label{eq:magicL2b}
\end{align}
After a Baikov parametrization and a maximal cut, we may solve $d \log(\mathcal{B}) =0$ and we get the two conjugate point-like solutions
\begin{align}
x_7 &= 2 m^2 \,, &  x_8 &= \big( 1 \mp \sqrt{5} \big) m^2 \,, & x_9 &= \big( 1 \pm \sqrt{5} \big) m^2 \,, & x_{10} &= \tfrac{1}{2} \big( 3 \mp \sqrt{5} \big) m^2 \,, \nonumber \\
x_{11} &= \tfrac{1}{2} \big( 3 \pm \sqrt{5} \big) m^2 \,, & x_{12} &= m^2 \,, & x_{13} &= \tfrac{1}{2} \big( 3 \mp \sqrt{5} \big) m^2 \,, & x_{14} &= \tfrac{1}{2} \big( 3 \pm \sqrt{5} \big) m^2 \,.
\end{align}
and the one-dimensional solution(s)
\begin{align}
x_7 &= 0\,,\quad\;\; x_8 = -\tfrac{4}{3} m^2\,,\quad\;\; x_9 = -\tfrac{4}{3} m^2\,, & x_{11} &= x_{10} \,,\quad\;\; x_{12} = 2 m^2 \,, \nonumber \\
x_{13} &= \tfrac{1}{3} \left( 4 m^2 \mp \sqrt{4 m^4 {+} 24 m^2 x_{10} {-} 9 x_{10}^2} \right) \,, & x_{14} &= \tfrac{1}{3} \left( 4 m^2 \mp \sqrt{4 m^4 {+} 24 m^2 x_{10} {-} 9 x_{10}^2} \right) \,,\;\;
\end{align}
(parametrized in terms of $x_{10}$) that we again see to belong to the same variety.

That variety is generated by
\begin{align}
\tilde{\mathcal{P}} = 4 m^4 - 8 m^2 x_{10} + 3 x_{10}^2 - 8 m^2 x_{13} + 3 x_{13}^2
\end{align}
which has one critical point
\begin{align}
x_{10} = x_{13} = \frac{4 m^2}{3}
\end{align}
leading us to conclude that the generating sector of $L_2$ has three master integrals (before the application of symmetries) in agreement with FIRE.

\subsubsection*{Example $L_3$} 
The next example is in ref.~\cite{Lee:2013hzt} known as $\#414$. We use a parametrization with the propagators
\begin{align}
P_1 &= k_3^2 \,, & P_2 &= k_4^2 \,, & P_3 &= k_1^2 - m^2 \,, \nonumber \\
P_4 &= (k_1 + k_4)^2 - m^2 \,, & P_5 &= k_2^2 - m^2 \,, & P_6 &= (k_3 + k_4 + p)^2 - m^2 \,;
\end{align}
and the ISPs
\begin{align}
P_7 &= (k_1 + k_2)^2\,, & P_8 &= (k_1 + k_3)^2\,, & P_9 &= (k_2 + k_3)^2\,, & P_{10} &= (k_2 + k_4)^2\,, \nonumber \\
P_{11} &= (k_3 + k_4)^2\,, & P_{12} &= (k_1 + p)^2\,, & P_{13} &= (k_2 + p)^2\,, & P_{14} &= (k_3 + p)^2\,.
\end{align}
The kinematics is such that $p^2 = m^2$.
The magic relation of $L_3$ is
\begin{align}
\adjustbox{raise=-2.4mm}{\includegraphics[width=0.10\textwidth]{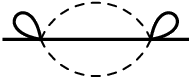}} \, - \,
\adjustbox{raise=-2.4mm}{\includegraphics[width=0.10\textwidth]{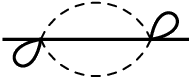}} &= 0
\end{align}
which is equivalent to a symmetry relation.

On the maximal cut we get one line-like solution for the critical point equations:
\begin{align}
x_7 &= 2 m^2\,, & x_9 &= m^2 \,, & x_{10} &= m^2\,, &
x_{11} &= \tfrac{8}{3} m^2 \,, \qquad \nonumber \\
x_{12} &= \tfrac{1}{2} ( 5 m^2 - x_8 ) \,, & x_{13} &= 2 m^2 \,, & x_{14} &= -\tfrac{1}{3} m^2 \,. &&
\end{align}
parametrized in terms of $x_8$. This straight line has no critical points itself, so we conclude that $L_3$ has no master integrals in its generating sector.

\subsubsection*{Example $L_4$} 
This example is in ref.~\cite{Lee:2013hzt} known as $\#429$. We use a parametrization with the propagators
\begin{align}
P_1 &= k_3^2 \,, & P_2 &= k_4^2 \,, & P_3 &= k_1^2 - m^2 \,, \nonumber \\
P_4 &= (k_1 + k_3 + k_4)^2 - m^2 \,, & P_5 &= k_2^2 - m^2 \,, & P_6 &= (k_4 + p)^2 - m^2 \,,
\end{align}
and the ISPs
\begin{align}
P_7 &= (k_1 + k_2)^2 \,, & P_8 &= (k_1 + k_4)^2 \,, & P_9 &= (k_2 + k_3)^2 \,, & P_{10} &= (k_2 + k_4)^2 \,, \nonumber \\
P_{11} &= (k_3 + k_4)^2 \,, & P_{12} &= (k_1 + p)^2 \,, & P_{13} &= (k_2 + p)^2 \,, & P_{14} &= (k_3 + p)^2 \,.
\end{align}
The kinematics is such that $p^2 = m^2$.
The magic relation of sector $L_4$ is
\begin{align}
2 (5-2d) \, \adjustbox{raise=-4mm}{\includegraphics[width=0.10\textwidth]{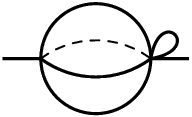}} \, + \, 8 m^2
\adjustbox{raise=-4mm}{\includegraphics[width=0.10\textwidth]{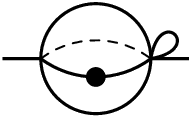}} \, + \, (d-2) \,
\adjustbox{raise=-4mm}{\includegraphics[width=0.075\textwidth]{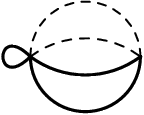}} \, &= \, 0
\end{align}
which after factorizing the one-loop tadpole out becomes
\begin{align}
\adjustbox{raise=-4mm}{\includegraphics[width=0.10\textwidth]{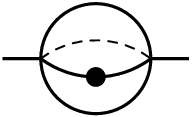}} \, &= \, \frac{2d-5}{4 m^2} \,
\adjustbox{raise=-4mm}{\includegraphics[width=0.10\textwidth]{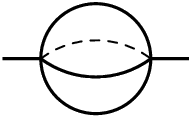}} \, + \, \frac{2-d}{8 m^2} \,
\adjustbox{raise=-4mm}{\includegraphics[width=0.06\textwidth]{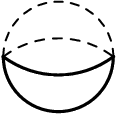}}
\end{align}
that we realize to be equivalent to eq.~\eqref{eq:sm1gr}.

On the maximal cut we get one line-like solution for $d \log( \mathcal{B}) = 0$:
\begin{align}
x_7 &= 2 m^2 \,, & x_8 &= -\tfrac{1}{3} m^2 \,, & x_9 &= m^2 \,, & x_{10} &= m^2 \,, \qquad \nonumber \\
x_{11} &= \tfrac{8}{3} m^2 \,, & x_{12} &= \tfrac{1}{2} ( 5 m^2 - x_{14} ) \,, & x_{13} &= 2 m^2 \,. &&
\end{align}
parametrized in terms of $x_{14}$, and again this straight line has no critical points, so we conclude there to be no master integrals in the generating sector of $L_4$.

\subsubsection*{Example $L_5$} 
The fifth example is in ref.~\cite{Lee:2013hzt} known as $\#821$. We use a parametrization with the propagators
\begin{align}
P_1 &= k_2^2 \,, & P_2 &= k_3^2 \,, & P_3 &= k_1^2 - m^2 \,, \nonumber \\
P_4 &= (k_1 + k_3)^2 - m^2 \,, & P_5 &= (k_2 + k_3 + k_4)^2 - m^2 \,, & P_6 &= k_4^2 - m^2 \,,
\end{align}
and the ISPs
\begin{align}
P_7 &= (k_1 + k_2)^2 \,, & P_8 &= (k_1 + k_4)^2 \,, & P_9 &= (k_2 + k_3)^2 \,, & P_{10} &= (k_2 + k_4)^2 \,.
\end{align}
Note that there are fewer ISPs here than in the other $L_i$ cases, since there is no external momentum.
The magic relation generated by $L_5$ is
\begin{align}
2 (5-2d) 
\adjustbox{raise=-4mm}{\includegraphics[width=0.06\textwidth]{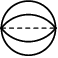}} + 8 m^2
\adjustbox{raise=-4mm}{\includegraphics[width=0.06\textwidth]{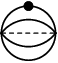}} + (d-2) \Bigg( 
\adjustbox{raise=-4mm}{\includegraphics[width=0.075\textwidth]{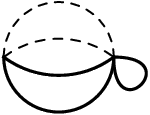}} -
\adjustbox{raise=-4mm}{\includegraphics[width=0.075\textwidth]{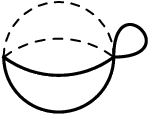}} \Bigg) = 0
\end{align}
which after imposing symmetries between the sectors becomes equivalent to eq.~\eqref{eq:magicL2b}.

We get, on the maximal cut, one line-like solution for $d \log( \mathcal{B}) = 0$ given by
\begin{align}
x_8 &= \tfrac{1}{2} (5 m^2 - x_7) \,, & x_9 &= \tfrac{8}{3} m^2 \,, & x_{10} = - \tfrac{1}{3} m^2 \,.
\end{align}
parametrized in terms of $x_7$, and again such a straight line solution indicates there to be no master integrals in the generating sector.

\subsubsection*{Example $L_6$} 
This next example is in ref.~\cite{Lee:2013hzt} known as $\#924$. We use a parametrization with the propagators
\begin{align}
P_1 &= k_2^2 \,, & P_2 &= k_3^2 \,, & P_3 &= k_4^2 \,, \nonumber \\
P_4 &= k_1^2 - m^2 \,, & P_5 &= (k_1 + k_4)^2 - m^2 \,, & P_6 &= (k_2 + k_3 + k_4 + p)^2 - m^2 \,;
\end{align}
and the ISPs
\begin{align}
P_7 &= (k_1 + k_2)^2 \,, & P_8 &= (k_1 + k_3)^2 \,, & P_9 &= (k_2 + k_3)^2 \,, & P_{10} &= (k_2 + k_4)^2 \,, \nonumber \\
P_{11} &= (k_3 + k_4)^2 \,, & P_{12} &= (k_1 + p)^2 \,, & P_{13} &= (k_2 + p)^2 \,, & P_{14} &= (k_3 + p)^2 \,.
\end{align}
The kinematics is such that $p^2 = m^2$.
The magic relation generated by sector $L_6$ is
\begin{align}
\adjustbox{raise=-4mm}{\includegraphics[width=0.10\textwidth]{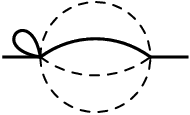}} \, - \,
\adjustbox{raise=-4mm}{\includegraphics[width=0.10\textwidth]{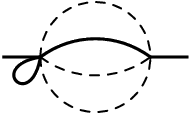}} &= 0
\end{align}
which is equivalent to a symmetry relation.
For the critical point analysis on the maximal cut we get one line-like solution
\begin{align}
x_8 &= x_7 \,, & x_9 &= m^2 \,, & x_{10} &= m^2 \,, & x_{11} &= m^2 \,, \qquad \nonumber \\
x_{12} &= 3 m^2 - x_7 \,, & x_{13} &= 0 \,, & x_{14} &= 0 \,. &&
\end{align}
parametrized in terms of $x_7$ and again a straight line implies no master integrals in the generating sector.

\subsubsection*{Example $L_7$} 
The final $L_i$ example is in ref.~\cite{Lee:2013hzt} known as $\#969$.  We use a parametrization with the propagators
\begin{align}
P_1 &= k_2^2 \,, & P_2 &= k_3^2 \,, & P_3 &= k_4^2 \,, \nonumber \\ 
P_4 &= k_1^2 - m^2 \,, & P_5 &= (k_1 + k_2 + k_3 + k_4)^2 - m^2 \,, & P_6 &= (k_4 + p)^2 - m^2 \,,
\end{align}
and the ISPs
\begin{align}
P_7 &= (k_1 + k_2)^2 \,, & P_8 &= (k_1 + k_3)^2 \,, & P_9 &= (k_2 + k_3)^2 \,, & P_{10} &= (k_2 + k_4)^2 \,, \nonumber \\
P_{11} &= (k_3 + k_4)^2 \,, & P_{12} &= (k_1 + p)^2 \,, & P_{13} &= (k_2 + p)^2 \,, & P_{14} &= (k_3 + p)^2 \,.
\end{align}
The kinematics is such that $p^2 = m^2$.
The magic relation generated by sector $L_7$ is
\begin{align}
(12-5d) \,
\adjustbox{raise=-4mm}{\includegraphics[width=0.10\textwidth]{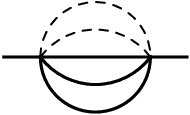}} \, + \, 8 m^2 \,
\adjustbox{raise=-4mm}{\includegraphics[width=0.10\textwidth]{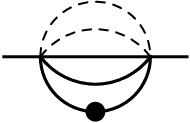}} \, + \, (d-2) \,
\adjustbox{raise=-4mm}{\includegraphics[width=0.06\textwidth]{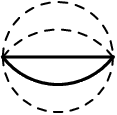}} &= 0
\label{eq:magic969}
\end{align}
which is unchanged by the application of symmetries.
After the critical point analysis we get one line-like solution
\begin{align}
x_7 &= 0 \,, & x_8 &= 0 \,, & x_9 &= m^2 \,, & x_{10} &= m^2 \,, \qquad \nonumber \\
x_{11} &= m^2 \,, & x_{12} &= 3 m^2 - x_{13} \,, & x_{14} &= x_{13} \,. &&
\end{align}
parametrized in terms of $x_{13}$. Again a straight line corresponds to there being no master integrals in the generating sector.

\subsection{Extra examples}
\label{sec:examplesextra}

Finally, we have two additional examples that we will refer to as $E_1$ and $E_2$, and in addition we will rephrase the Higgs example from the main text in terms of the Baikov representation.

\subsubsection*{Example $E_1$} 
This example is inspired by the example files of the Landau analysis program SOFIA~\cite{Correia:2025wtb}. There is\footnote{We thank Mathieu Giroux and Sebastian Mizera for discussions about this example.} a case, there called the ``degenerate acnode'', that is surprisingly similar to our Higgs example from \cref{sec:magic}, and indeed it turns out to also have a magic relation.

We use the parametrization
\begin{align}
P_1 &= k_1^2 - m^2\,, & P_2 &= (k_1-p_1)^2 - m^2\,, & P_3 &= (k_2 - p_1 - p_2)^2 - m^2\,, \nonumber \\
P_4 &= (k_2 + p_4)^2 - m^2\,, & P_5 &= (k_1-k_2)^2 - m^2 \; ; \!\! & P_6 &= (k_1-p_1-p_2)^2 - m^2\,, \\
P_7 &= k_2^2 - m^2\,, & P_8 &= (k_1+p_4)^2 - m^2\,, & P_9 &= (k_2-p_1)^2 - m^2\,, \nonumber
\end{align}
where $P_1-P_5$ are the propagators and $P_6-P_9$ are the ISPs needed for standard Baikov. The kinematics is
\begin{align}
p_1^2 = p_2^2 = p_3^2 = p_4^2 = 0 \,, \qquad (p_1+p_2)^2 = s \,, \qquad (p_2+p_3)^2 = t \,.
\end{align}
The magic relation is
\begin{align}
s ( I_{11201} - I_{20111} ) + t ( I_{11021} - I_{02111} ) &= 0
\end{align}
or graphically
\begin{align} s \times \Bigg( \adjustbox{raise=-5mm}{\includegraphics[height=12.0mm]{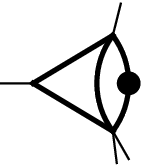}} \; - \; \adjustbox{raise=-5mm}{\includegraphics[height=12.0mm]{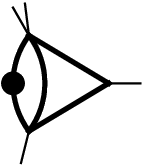}} \Bigg) \; + \; t \times \Bigg( \adjustbox{raise=-5mm}{\includegraphics[height=12.0mm]{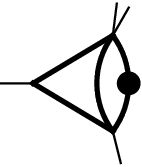}} \; - \;  \adjustbox{raise=-5mm}{\includegraphics[height=12.0mm]{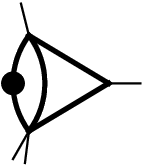}} \Bigg) \; = \; 0
\end{align}
which becomes trivial after the use of symmetries.

Solving the critical equations on the maximal cut gives five distinct solutions. Four of them are point-like
\vspace{-2mm}
\begin{align}
x_6 = x_7 = x_8 = x_9 = - m^2 \,; \qquad\quad x_6 = x_7 = x_8 = x_9 = \frac{st}{3 (s+t)} \,; \nonumber
\end{align}
\vspace{-8mm}
\begin{align}
x_6 = x_8 = - m^2\;, \quad x_7 = x_9 = 2 m^2 + \frac{st}{s+t} \,;
\end{align}
\vspace{-8mm}
\begin{align}
x_6 = x_8 = 2 m^2 + \frac{st}{s+t} \;, \quad x_7 = x_9 = - m^2 \,; \nonumber
\end{align}
but the last is the one-dimensional
\begin{align}
x_6 &= \frac{s (t-x_8)}{t}\,, & x_7 &= \frac{- s t (m^2 + x_8)}{st - (s+t) x_8}\,, & x_9 &= \frac{t \big( t (m^2 + s) - s x_8 \big)}{st - (s+t) x_8}\,,
\end{align}
parametrized by $x_8$. Its non-trivial generator is
\begin{align}
\tilde{\mathcal{P}} &= t^2 (m^2 + s) - st (x_8 + x_9) + (s + t) x_8 x_9
\end{align}
which has one critical point
\begin{align}
x_8 = x_9 = \frac{st}{s+t}
\end{align}
so we have five critical points in total, and FIRE agrees that the generating sector of $E_1$ has five master integrals (before applying symmetries).

\begin{figure}
\centering
\adjustbox{raise=0mm}{\includegraphics[width=0.21\textwidth]{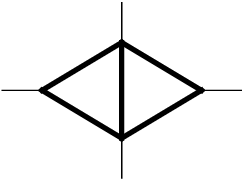}} \hspace{12mm}
\adjustbox{raise=0mm}{\includegraphics[width=0.21\textwidth]{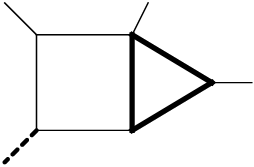}}
\caption{Generating sectors for examples $E_1$ and $E_2$. Thin lines are massless, thick lines have squared mass $m^2$, and the thick dotted line has squared mass $\mu_4$.}
\label{fig:Esectors}
\end{figure}

\subsubsection*{Example $E_2$} 
This example comes from the example files of the IBP program Kira\footnote{We thank Johann Usovitsch for discussions on this example.}~\cite{Maierhofer:2018gpa, Lange:2025fba}, but we have here changed the notation and conventions\footnote{In the actual example from the Kira example files, $p_2^2 \neq 0$ and in addition $p_2^2$ and $p_4^2$ are put to specific numerical values. That is not needed for the example to work.}. We will use the parametrization
\begin{align}
P_1 &= k_1^2 \,, & P_2 &= (k_1+p_1)^2 \,, & P_3 &= (k_1-p_2-p_3)^2 \,, \nonumber \\
P_4 &= (k_2-p_2-p_3)^2 - m^2 \,, & P_5 &= (k_2-p_2)^2 - m^2 \,, & P_6 &= (k_1-k_2)^2 - m^2 \,,
\label{eq:propagatorsE2}
\end{align}
and we need 3 ISPs which we choose as
\begin{align}
P_7 &= (k_1-p_2)^2 \,, & P_8 &= k_2^2 - m^2 \,, & P_9 &= (k_2+p_1)^2 - m^2 \,.
\end{align}
The kinematics is such that
\begin{align}
p_1^2 = p_2^2 = p_3^2 = 0 \,, \quad p_4^2 = \mu_4 \,, \quad (p_1+p_2)^2 = s \,, \quad (p_2+p_3)^2 = t \,.
\end{align}

The magic relation generated by sector $E_2$ is
\begin{align}
I_{110122} + I_{110212} - 2 I_{110113} + 2 I_{111013} - 2 I_{111031} &= 0
\end{align}
or graphically
\begin{align}
\adjustbox{raise=-6.5mm}{\includegraphics[width=0.12\textwidth]{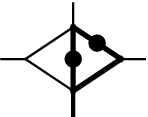}} \, + \, 
\adjustbox{raise=-6.5mm}{\includegraphics[width=0.12\textwidth]{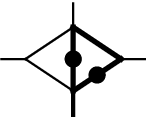}} \, - \, 2 \times 
\adjustbox{raise=-6.5mm}{\includegraphics[width=0.12\textwidth]{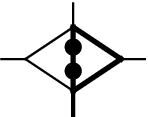}} \, + \, 2 \times \!\!\!
\adjustbox{raise=-6.5mm}{\includegraphics[width=0.09\textwidth]{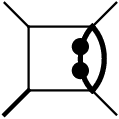}} \, - \, 2 \times \!\!\!
\adjustbox{raise=-6.5mm}{\includegraphics[width=0.09\textwidth]{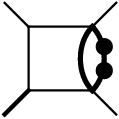}}
 = 0
\end{align}
Due to a symmetry relation in the $I_{111011}$-sector the last two terms cancel, leaving
\begin{align}
I_{110122} + I_{110212} - 2 I_{110113} &= 0,
\label{eq:magicE2c}
\end{align}
looking like an extra relation\footnote{In the Kira example files there is a second family defined as in eq.~\eqref{eq:propagatorsE2} except that $P_3$ is replaced by $P_3' = (k_1{-}k_2{+}p_2)^2 - m^2$ making the family non-planar. That family has no magic relations, but it still has the three integrals of eq.~\eqref{eq:magicE2c} as subsectors, making how to extract eq.~\eqref{eq:magicE2c}, while working entirely within that non-planar family, an open question. \label{footnote:kiraextra}} in the $I_{110111}$-sector.

We may then do the standard Baikov parametrization and a maximal cut leaving three uncut variables. In those we will solve $d \log( \mathcal{B}) = 0$, giving a set of conjugate one-dimensional solutions corresponding to
\begin{align}
x_7 = 0 \;, \;\;\; x_8 = \frac{\, t \big( s \mu_4 - (2 \mu_4 {-} s) x_9 \big) \pm \sqrt{st \big( st (\mu_4{-}x_9)^2 - 8 m^2 \mu_4 (\mu_4{-}s)(\mu_4{-}t) \big) \,}}{2 \mu_4 (s - \mu_4)}
\end{align}
Together the two solution branches form a hyperbola.
The non-trivial generator is
\begin{align}
\tilde{\mathcal{P}} &= 2 m^2 (\mu_4 {-} t) s t + (\mu_4 x_8 - t x_9) \big( s t + (\mu_4 {-} s) x_8 - t x_9 \big)
\end{align}
which itself has one critical point given by
\begin{align}
x_8 = t \,, \qquad x_9 = \mu_4 \,,
\end{align}
allowing us to conclude that there is one master integral in the generating sector of $E_2$ in agreement with Kira.

\subsubsection*{Example $H$ -- The Higgs example in Baikov representation}

As our last example, let us revisit the Higgs example from sections \ref{sec:magic}, \ref{sec:criticalvarieties} and \ref{sec:Higgspart3}. Where its critical varieties in \cref{sec:criticalvarieties} were discussed in terms of the Lee--Pomeransky representation, we will here rephrase that discussion in terms of the Baikov representation. The propagators, including the four ISPs needed for the standard Baikov representation, are given by eq.~\eqref{eq:higgspara} and the kinematics is given by eq.~\eqref{eq:higgskinematics}. The magic relation generated from the five-propagator sector depicted on \cref{fig:higgsgen}, is discussed in \cref{sec:magic} and given by eq.~\eqref{eq:magichiggs}.

To investigate the critical varieties, one can perform a Baikov parametrization, do the maximal cut, and build the Jacobian ideal as described in \cref{sec:numberofmasters}:
\begin{align}
\mathcal{I} = \big\langle \partial_{x_6} {\mathcal{B}} \,,\; \partial_{x_7} {\mathcal{B}} \,,\; \partial_{x_8} {\mathcal{B}} \,,\; \partial_{x_9}{\mathcal{B}} \,,\; 1 - x_0 {\mathcal{B}} \big\rangle
\end{align}
Doing the primary decomposition allows for a formulation in terms of the intersection of three primary ideals, as was the case in the Lee--Pomeransky representation in \cref{sec:criticalvarieties}. The result is
\begin{align}
\mathcal{I}_1 &= \Big\langle F + 3 u x_6 \,,\;\;\,  F + 3 u x_7 \,,\;\;\, F + 3 u x_8 \,,\;\;\, F + 3 u x_9 \,,\;\;\, 27 u - F^3 x_0 \Big\rangle \\
\mathcal{I}_2 &= \Big\langle st + u (x_8+x_9) \,,\;\;\, x_7-x_9 \,,\;\;\, st + u (x_6+x_9) \,, \nonumber \\[-0.5mm]
& \qquad\quad u x_9^2 + st x_9 + 2 m_t^2 st \,,\;\;\, 1 - m_t^2 s^2 t^2 x_0 \Big\rangle \\
\mathcal{I}_3 &= \Big\langle st - t x_7 + (m_H^2{-}s) x_9 \,,\;\;\, s t + (m_H^2{-}t) x_6 - s x_8 \,, \;\;\, \tilde{\mathcal{P}} ,\;\;\, 1 - 4 m_t^4 s t u x_0 \Big\rangle
\end{align}
where we have defined
\begin{align}
u := m_H^2 {-} s {-} t \,, \qquad\;\;\; F := 4 m_t^2 u + st \,,
\end{align}
and
\begin{align}
\tilde{\mathcal{P}} := -t \big( st - 2 m_t^2 (m_H^2 {-} t) \big) + st (x_8 + x_9) + u x_8 x_9
\label{eq:Ptildehiggsbaikov}
\end{align}

Computing the critical varieties, we get one critical point from $\mathcal{I}_1$
\begin{align}
x_6 &= x_7 = x_8 = x_9 = - \frac{4 m_t^2 u + st}{3 u}
\end{align}
two conjugate critical points from $\mathcal{I}_2$:
\begin{align}
x_6 &= x_8 = - \frac{s t \pm \sqrt{s t ( s t - 8 m_t^2 u) }}{2 u} \;,\;\; \quad x_7 = x_9 = - \frac{s t \mp \sqrt{s t ( s t - 8 m_t^2 u)}}{2 u} \;,
\end{align}
and from $\mathcal{I}_3$ we get the one-dimensional critical variety:
\begin{align}
x_6 = \frac{-st (2 m_t^2 + x_9)}{st + u x_9}\;, \quad x_7 = \frac{st + (m_H^2{-}s) x_9}{t}\;, \quad x_8 = \frac{t ( st + 2 m_t^2 (t {-} m_H^2) - s x_9)}{st + u x_9}\;,
\end{align}
parametrized in terms of $x_9$. We will then proceed with looking for critical points \textit{on} the critical variety, in accordance with the full Lee--Pomeransky criterion of \cref{sec:criticalvarieties}, by finding critical points of the $\tilde{\mathcal{P}}$ as given by eq.~\eqref{eq:Ptildehiggsbaikov}. We get the point
\begin{align}
x_8 = x_9 = \frac{-st}{u}
\end{align}
so in total we find four critical points, telling us that there are four master integrals in the generating sector, as it also was found in \cref{sec:criticalvarieties}.

We note that the structure of the primary ideal decomposition is similar in the two representations, in that there are three primary ideals, where one has one critical point, one has two conjugate critical points, and the last has a one-dimensional critical variety, itself with one critical point.

\subsection{Summary of the examples}

In each of the examples in this section
we found a magic relation along with a higher dimensional critical variety. We will here attempt to summarize their properties.

As discussed in \cref{sec:magicandsymmetry} magic relations can, after the use of symmetry relations, be sorted into three symmetry types. Type A is those that keep relating integrals in different subsectors, type B is those that relate integrals in one subsector, and type C is those that trivialize. Our above examples have cases of each:

Type A: $\;\{H, S_1, S_2, L_4, L_7\}$,

Type B: $\;\{L_2, L_5, E_2\}$,

Type C: $\;\{T_1, T_2, L_1, L_3, L_6, E_1\}$.

Similarly we can categorize the higher dimensional varieties that we find after their dimensionalities and their own number of critical points, which, according to the full Lee--Pomeransky criterion of \cref{sec:criticalvarieties}, determines how they contribute to the counting of master integrals in the generating sector.

Examples $\{T_1, S_1, S_2, L_3, L_4, L_5, L_6, L_7\}$ have a critical variety that is a (one-dimensional) straight line which does not contribute to the master integral counting in the generating sector.

Examples $\{H, L_1, L_2, E_1, E_2\}$ have a critical variety that is a (one-dimensional) conic, which contributes one to the master integral counting in the generating sector.

Example $T_2$ has a critical variety that is a two-dimensional plane which does not contribute to the master integral counting in the generating sector.

Based on the admittedly low statistics of these 14 examples, magic relations appear to be rather equally distributed between the three symmetry types. Regarding the higher dimensional critical varieties on the other hand, all but one are one-dimensional, and our only two-dimensional example was explicitly constructed to have that property. Also we found no critical varieties of polynomial degree higher than two (corresponding to a conic). We note that we do not observe, nor did we expect, any correlation between the properties of the critical variety and the symmetry type of the magic relations.

\section{Discussion and future prospects}
\label{sec:discussion}
In this work, we observed that \hyperlink{box:magicDef}{magic relations}---integration-by-parts identities for which the generating sector decouples---are closely linked to the existence of higher-dimensional components in the critical variety of the twist, namely the locus where
$\omega := \dd\log\Phi = 0$. Specifically, we conjectured that a sector produces a magic relation if and only if the critical variety of that sector has a higher-dimensional component.

We supported this conjecture via a connection to Koszul cohomology and by arguing, under certain assumptions, that magic relations cannot be generated when the critical variety is zero-dimensional (\cref{sec:proof} and appendix \ref{app:trivialargument}). 
We also presented numerous explicit examples (\cref{sec:examples}) that support our conjecture. 

The procedure to correctly count master integrals in the presence of higher-dimensional critical varieties, known as the full Lee--Pomeransky criterion, was also discussed in detail. This criterion was given in ref.~\cite{Lee:2013hzt} but does not appear to be widely known. We link the full Lee--Pomeransky criterion to arguments from Morse--Bott theory in appendix~\ref{app:leepomandmorse}.
Finally we categorized the ways in which magic relations interact with symmetry relations.

Beyond a theoretical understanding, the results of this work have important practical implications for the computation of Feynman integrals. In particular they allow for an efficient way to test for the existence of magic relations, without having to find them explicitly by building and solving large IBP systems. Furthermore, our observations show that we may identify magic relations as non-trivial divergence-free syzygies, allowing us to find them systematically. These ideas were implemented in simple and efficient proof of concept \textsc{Mathematica} functions \texttt{MagicQ} and \texttt{FindMagicRelations} provided in the ancillary file \texttt{Magic-Test.m}.

Despite the progress made in this work, there are still many important unresolved questions regarding magic relations. Perhaps the most important problem concerns the interplay of magic relations and cuts, discussed in \cref{sec:magicandcuts}. Previously, it has been assumed that treating cuts as a residue operation on the integrand will leave linear equations generated from IBPs invariant. For the case of magic relations, this is clearly no longer true. The reason for \textit{why} this almost universal procedure breaks down is however unclear, and remains an important gap in the current theoretical understanding of these IBPs. This motivates the need for a new prescription of how to treat cuts in IBP algorithms. Resolving this issue could presumably also lead to a resolution of how to correctly compute relative twisted cohomology intersection numbers in such cases, which is currently poorly understood.

Another interesting unresolved question concerns magic relations of ``type B'', namely relations which, after the application of symmetry relations, take the same form as an IBP identity generated from a single subsector of the generating sector.
This suggests that additional IBP-like identities for a given integral family may be obtained by considering magic relations generated from super-sectors, corresponding to embeddings of the family into a larger one\footnote{We discuss an example of this in \cref{footnote:kiraextra} in the context of example $E_2$.}.
Such additional identities could simplify IBP reduction and reduce the number of master integrals. Nevertheless it is unclear if there exists a systematic way of finding such embeddings.

The critical variety is also an important ingredient 
in the study of singularities of Feynman integrals after integration~\cite{Landau:1959fi}.
The Landau locus corresponds to the region in kinematic parameter space where the topology of the critical variety changes discontinuously. This suggests that Feynman integrals with higher-dimensional critical varieties, such as those studied in this manuscript, may exhibit Landau singularities with unusual behavior.
This possibility is supported by the fact that example $E_1$ was taken from the Landau analysis program SOFIA~\cite{Correia:2025wtb}, where it was used as an example of contour degeneracy. A preliminary analysis suggests that similar behavior occurs for some, but not all, of our other examples.
The connection between higher-dimensional critical varieties and Landau singularities therefore remains an interesting direction for future work.

Finally, proving completely the \hyperlink{box:keyObs}{Key observation} is a clear avenue for further exploration, and could provide a deeper understanding and explanations for the unresolved questions mentioned above.

\subsection*{Acknowledgements}

We would like to thank Wojciech Flieger and Felix Tellander for many enlightening conversations. We furthermore thank Claude Duhr, Federico Gasparotto, Mathieu Giroux, Martin Helmer, Franz Herzog, Luke Lippstreu, Pierpaolo Mastrolia, Andrew McLeod, Sebastian Mizera, Maria Polackova, Johann Usovitsch and Li Lin Yang for many extremely helpful discussions. We would also like to thank Ruth Britto, Claude Duhr, Yuhang Jia, Pierpaolo Mastrolia, Cathrin Semper, Stefan Weinzierl, Mao Zeng, and Yang Zhang, for providing comments on the manuscript in its preliminary forms. 

GC’s research is supported by the United Kingdom Research and Innovation grant UKRI FLF MR/Y003829/1.
HF was supported in part by the Excellent Young Scientists Fund Program of the National Natural Science Foundation of China (NSFC).
AP is supported by the European Union (ERC, UNIVERSE PLUS, 101118787). S.S. research is partially supported by the Amplitudes INFN scientific initiative.

\appendix

\section{The cut of the Higgs parachute}\label{app:cut_calc}

In this appendix we will evaluate the maximal cut of the integral $I_{11101}$ discussed in eq.~\eqref{eq:magiconcut}. We will do this using the loop-by-loop Baikov representation~\cite{Frellesvig:2017aai, Frellesvig:2024ymq}, which implies that not all the propagators listed in eq.~\eqref{eq:higgspara} are needed. We will need only
\begin{align}
P_1 &= k_1^2 - m_t^2\,, & P_2 &= (k_1{-}p_1)^2 - m_t^2\,, & P_3 &= (k_2{-}p_1{-}p_2)^2 - m_t^2\,, \nonumber \\
P_5 &= (k_1{-}k_2)^2\,; & P_6 &= (k_1{-}p_1{-}p_2)^2 - m_t^2 \,, &&
\end{align}
and the kinematics is as given by eq.~\eqref{eq:higgskinematics}, specifically
\begin{align}
p_1^2 = p_2^2 = 0\,, \qquad (p_1{+}p_2)^2 = s\,.
\end{align}
With this, the loop-by-loop Baikov representation\footnote{One could consider doing the same calculation starting from the standard Baikov representation, using all nine of the propagator-type objects given by eq.~\eqref{eq:higgspara}. This can be done, but then the extra Baikov variables $\{x_4,x_7,x_8,x_9\}$ can be integrated out one by one, as explained e.g. in refs.~\cite{Jiang:2023qnl, Frellesvig:2024ymq}, giving the same expression as eq.~\eqref{eq:HiggsLBL}. These integrations have to be done in an order where each step is compatible with a loop-by-loop representation. For instance the order $x_9, x_7, x_4, x_8$ is valid.} is
\begin{align}
I_{a_1 a_2 a_3 0 a_5} = K \mathcal{E}_2^{(3{-}d)/2} \int \frac{\mathcal{B}_2^{(d{-}4)/2} \, \mathcal{E}_1^{(2{-}d)/2} \, \mathcal{B}_1^{(d{-}3)/2} \, d^5 x}{x_1^{a_1} x_2^{a_2} x_3^{a_3} x_5^{a_5}}
\,,
\label{eq:HiggsLBL}
\end{align}
where
\begin{align}
K = \frac{1}{8 \, \pi^{3/2} \, \Gamma((d{-}2)/2) \, \Gamma((d{-}1)/2)}
\,,
\end{align}
and
\begin{eqnarray}
\mathcal{E}_2 \, = & G(p_1, p_2) \;\;\;\;\;\;\;\;\;\;\;\; \;
& = \, - \tfrac14 s^2 \,,\\
\mathcal{B}_2 \, = & G(k_1, p_1, p_2) \;\;\;\;\;\;\; \;
& = \, - \tfrac14 s \big( s (x_2{+}m_t^2) + (x_2-x_1)(x_2-x_6) \big) \,,\\
\mathcal{E}_1 \, = & G(k_1{-}p_1{-}p_2) \;\;\;\;\; \;
& = \, x_6 + m_t^2 \,, \\
\mathcal{B}_1 \, = & \, G(k_2, k_1{-}p_1{-}p_2) \, \;
& = \, - \tfrac14 \big( x_5^2 - 2 x_5 ( 2 m_t^2 {+} x_3 {+} x_6) + (x_3{-}x_6)^2 \big) \,,
\end{eqnarray}
where the $G$ denotes a Gram determinant.
We can then perform the maximal cut by acting with $\mathrm{res}_{x_i=0}$ for $i \in \{1,2,3,5\}$ and renaming $x_6 \rightarrow z$. After the cut, the integrand depends on the polynomials
\begin{align}
\mathcal{E}_2 &= - \tfrac14 s^2 \,,   &  \mathcal{B}_2 &= - \tfrac14 s^2 m_t^2 \,, & \mathcal{E}_1 &= z + m_t^2 \,, &  \mathcal{B}_1 &= - \tfrac14 z^2 \,.
\end{align}
Inserting the cut polynomials gives
\begin{align}
I_{11101}|_{\text{cut}} = \tilde{K} \int (z{+}m_t^2)^{(d-2)/2} \, z^{d-3}  \, \text{d} z \,,
\end{align}
where
\begin{align}
\tilde{K} = \frac{(-1)^{d/2} \, 2^{1-d} \, (m_t^2)^{(d-4)/2}}{\pi^{3/2} \, s \, \Gamma((d{-}2)/2) \, \Gamma((d{-}1)/2)} \,.
\end{align}
We see that there \textit{are} integration contours where the cut is zero. For instance the contour where $\mathcal{B}_1 \geq 0$ will, after the cut, be from $0$ to $0$ and thus give a vanishing result. However crucially there are also contours where the cut is non-zero. Importantly, IBP identities are contour independent and thus should hold for all (allowed) choices. Picking $-m_t^2$ to $0$, we get
\begin{align}
I_{11101}|_{\text{cut}} &= \tilde{K} \, \frac{(-1)^{d-1} \, (m_t^2)^{(d-2)/2} \, \Gamma ((4{-}d)/2) \, \Gamma(d{-}2)}{\Gamma(d/2)} \,,\\
&= \frac{(-1)^{3d/2} \, (m_t^2)^{d{-}3} \, \Gamma((4{-}d)/2)}{4 \pi^2 s \Gamma(d/2)} \, \neq \, 0\,.
\end{align}

\section{The full Lee--Pomeransky criterion and Morse--Bott theory}
\label{app:leepomandmorse}

In this appendix we discuss the Lee--Pomeransky criterion in its restricted and full forms. For the \textit{restricted} criterion we prove it following ref.~\cite{Weinzierl:2022eaz} (particularly exercises 62 and 63 at the end of section 6.7.2 therein). For the \textit{full} criterion we perform a parallel derivation within the framework of Morse--Bott theory, using information from ref.~\cite{BanyagaHurtubiseMorse}. Our discussion here builds on \cref{sec:numberofmasters} in which we described how the counting of Feynman integrals depends on the properties of $\log(|\Phi|)$ (as defined in eq.~\eqref{eq:multivalued}). If $\log(|\Phi|)$ is a Morse function, defined as a function with only isolated critical points, the number of master integrals is given by the restricted Lee--Pomeransky criterion of eq.~\eqref{eq:restrictedleepom}, whereas if the function has higher-dimensional critical varieties, the counting requires the full Lee--Pomeransky criterion of eq.~\eqref{eq:fullleepom}.

\subsection{The Morse case and the restricted Lee--Pomeransky criterion}
\label{app:morse}

For a Morse function we have the \textit{Morse inequalities}~\cite{Weinzierl:2022eaz, BanyagaHurtubiseMorse}
\begin{align}
\sum_{k=0}^{a} (-1)^{a-k} b_k \, \leq \, \sum_{k=0}^{a} (-1)^{a-k} m_k
\label{eq:morseineq}
\end{align}
with one inequality for each value of $a \in \mathbb{N}_0$. Here $b_k$ are the \textit{Betti numbers}
\begin{align}
b_k := \text{dim}(H_k)
\end{align}
and $m_k$ is the number of critical points with Morse index $k$, where the Morse index is the number of \textit{downward directions} from the critical point.
When we operate in $\mathbb{C}^n$ there are as many upward directions as downward directions from each point, so $m_k |_{k{\neq}n} = 0$. Let us define $m := m_n$, giving in total for the $m$s
\begin{align}
m_k = \delta_{k,n} \, m
\end{align}
which put together corresponds to
\begin{align}
\sum_{k=0}^{a} (-1)^{a-k} b_k \, \leq \, \sum_{k=0}^{a} (-1)^{a-k} \delta_{k,n} \, m
\label{eq:morseineq2}
\end{align}
Inserting the value $a=0$, and using the fact that the $b_k$ cannot be negative, we can easily derive $b_0=0$. Continuing inserting higher values of $a$ we likewise get that $b_k = 0$ for all $k$ smaller than $n$. The values $a=n$ and $a=n{+}1$ give us the relations
\begin{align}
b_{n} \leq m \;,\qquad b_{n{+}1} - b_n \leq - m 
\end{align}
which have as their only solution $b_n = m$ and $b_{n+1} = 0$, and continuing inserting even higher values of $a$ gives $b_k = 0$ for $k > n{+}1$. In total this corresponds to
\begin{align}
b_{k} = \delta_{k,n} \, m
\label{eq:leepomapp}
\end{align}
Given then the duality between homology and cohomology, this means that there are $m$ $n$-forms in the cohomology basis, and no $k$-forms with $k \neq n$, and that is the statement of the restricted Lee--Pomeransky criterion of eq.~\eqref{eq:restrictedleepom}.

\subsection{The Morse--Bott case and the full Lee--Pomeransky criterion}
\label{app:morsebott}

When a function instead of critical points has an assortment of higher dimensional critical varieties, and the space in which it is defined is \textit{compact}, it is known as a Morse--Bott function~\cite{BanyagaHurtubiseMorse}. For that case, instead of the Morse inequalities of eq.~\eqref{eq:morseineq} we have\footnote{This way of stating the Morse--Bott inequalities is not in ref.~\cite{BanyagaHurtubiseMorse}. It can be derived from theorem 3.53 of ref.~\cite{BanyagaHurtubiseMorse}, in the same way as the Morse inequalities of eq.~\eqref{eq:morseineq} (theorem 3.33 of ref.~\cite{BanyagaHurtubiseMorse}) is derived from theorem 3.36 of ref.~\cite{BanyagaHurtubiseMorse} in lemma 3.43 of ref.~\cite{BanyagaHurtubiseMorse}.}
the Morse--Bott inequalities:
\begin{align}
\sum_{k=0}^{a} (-1)^{a-k} b_k \; \leq \; \sum_{k=0}^{a} (-1)^{a-k} \, \sum_{i} b_{k-\lambda(\mathcal{C}_i)} (\mathcal{C}_i)
\label{eq:morsebott}
\end{align}
Here the $\mathcal{C}_i$ are the individual critical varieties, $\lambda(\mathcal{C}_i)$ is ``the number of unstable dimensions'' of $\mathcal{C}_i$, and $b_j(\mathcal{C}_i)$ are the Betti numbers of the varieties.
In $\mathbb{C}^n$ $\lambda(\mathcal{C}_i)$ corresponds to the co-dimension of the variety, so
\begin{align}
\lambda(\mathcal{C}_i) = n - d_i
\end{align}
where $d_i$ is the dimensionality of $\mathcal{C}_i$.

If all the critical varieties are points, then $\lambda(\mathcal{C}_i) = n$ and $b_{j} (\mathcal{C}_i) = \delta_{j,0}$ making $b_{k - \lambda(\mathcal{C}_i)} (\mathcal{C}_i) = \delta_{k,n}$. Then $\sum_{i} b_{k-\lambda(\mathcal{C}_i)} (\mathcal{C}_i) = \delta_{k,n} m$ and eq.~\eqref{eq:morsebott} reduces to the standard Morse case.

Using eq.~\eqref{eq:leepomapp} we get, for the individual critical varieties $\mathcal{C}_i$
\begin{align}
b_{j} (\mathcal{C}_i) = \delta_{j, d_i} m_{(i)}
\end{align}
where the $m_{(i)}$ are the numbers of critical points of the varieties. Putting this together gives
\begin{align}
b_{k-\lambda(\mathcal{C}_i)} (\mathcal{C}_i) \,=\, \delta_{k,n} m_{(i)}
\end{align}
which we now may insert into eq.~\eqref{eq:morsebott}
giving
\begin{align}
\sum_{k=0}^{a} (-1)^{a-k} b_k \; \leq \; \sum_{k=0}^{a} (-1)^{a-k} \delta_{k,n} M \qquad\;\; \text{where} \qquad\;\; M := \sum_{i} m_{(i)}
\end{align}
We see that this looks exactly like the standard Morse case of eq.~\eqref{eq:morseineq2} with $M$ playing the role of $m$. We can now go through all the same steps as above (from eq.~\eqref{eq:morseineq2} to eq.~\eqref{eq:leepomapp}), and would end up deriving
\begin{align}
b_{k} = \delta_{k,n} \sum_{i} m_{(i)}
\end{align}
Taking the liberty of promoting this result from the Morse--Bott framework to the framework of Feynman integrals, we can use the duality between homology and cohomology to reproduce the full Lee--Pomeransky criterion of eq.~\eqref{eq:fullleepom}.

\section{Trivial syzygies do not produce magic relations}\label{app:trivialargument}
In this appendix we will argue that trivial syzygies, as defined in~\cref{sec:proof}, cannot produce magic relations. For this argument we rely on the gcd condition in~\cref{eq:gcd_condition}
\begin{equation}\label{eq:gcd_condition_2}
    \partial_i\, \text{gcd}(\partial_{i}\mathcal{G},\partial_{j}\mathcal{G})=\partial_j\, \text{gcd}(\partial_{i}\mathcal{G},\partial_{j}\mathcal{G})=0\,.
\end{equation}
A basis for all trivial syzygies is given by~\cite{Lairez_2015}
\begin{align}\label{eq:div_free_basis}
    \phi_{i} = \partial_{j}\mathcal{G}, \quad \phi_{j} = -\partial_{i}\mathcal{G}, \quad \phi_{k}=0 \quad \text{for} \quad k\neq i,j\,.
\end{align}
We can see that these are included in~\cref{eq:KoszulExactnessInCoDeg2} by setting $\phi_{ij}=\mathcal{G}$ with all the other $\phi_{lm}=0$. These trivial syzygies also happen to be divergence free, however to span the full possible set of allowed syzygies these generators can be further multiplied by a polynomial $f$,
\begin{align}\label{eq:general_trivial_div_free_syzygy}
    \phi_{i} = f\,\partial_{j}\mathcal{G}\,, \quad \phi_{j} = -f\,\partial_{i}\mathcal{G}\,, \quad \phi_{k} = 0 \; \text{ for } \; k\neq i,j\,,
\end{align}
where $f$ must be chosen to still satisfy the divergence free condition in \cref{eq:div_free_cond}. For now we assume $f=1$. Let us now consider eq.~\eqref{eq:LP_Syz_IBP_2} evaluated on a basis element from eq.~\eqref{eq:div_free_basis}. We have
\begin{equation}\label{eq:IBP_trivial_top_sector}
    0=
    \int_{\partial_j \Gamma} \left[\mathcal{G}^{-d/2}\,\partial_i \mathcal{G}\right]_{x_j=0}\,  \dd \widehat{x}_j
    - 
    \int_{\partial_i \Gamma}\left[\mathcal{G}^{-d/2}\, \partial_j \mathcal{G}\right]_{x_i=0}\, \dd \widehat{x}_i\,.
\end{equation}
This identity, whilst generated from the top sector, can be reduced to zero by only considering IBP identities originating in the sectors one level below: taking the first term of \cref{eq:IBP_trivial_top_sector} one can write
\begin{align}
\int_{\partial_j \Gamma} \! \left[\mathcal{G}^{-d/2}\,\partial_{i} \mathcal{G} \right]_{x_j=0}\, \dd \widehat{x}_{j}
&=
\frac{2}{d-2}
\int_{\partial_j \Gamma} \! \dd \left(\left[\mathcal{G}^{-d/2}\,\mathcal{G} \right]_{x_j=0}\, \dd \widehat{x}_{ij}\right) \,.
\end{align}
Via Stokes' theorem (IBPs in the $a_j=0$ subsector) this can be rewritten as
\begin{align}
\int_{\partial_j \Gamma} \! \dd \left(\left[\mathcal{G}^{-d/2}\,\mathcal{G} \right]_{x_j=0}\, \dd \widehat{x}_{ij}\right) &= - \int_{\partial_{ij} \Gamma} \! \left[\mathcal{G}^{-d/2}\,\mathcal{G} \right]_{x_i=x_j=0} \dd \widehat{x}_{ij}\,.
\end{align}
giving in total
\begin{align}\label{eq:IBP_on_boundary}
\int_{\partial_j \Gamma} \! \left[\mathcal{G}^{-d/2}\,\partial_{i} \mathcal{G} \right]_{x_j=0}\, \dd \widehat{x}_{j} &= \frac{2}{d-2} \int_{\partial_{ij} \Gamma} \!\left[\mathcal{G}^{-d/2}\,\mathcal{G} \right]_{x_i=x_j=0} \dd \widehat{x}_{ij}\,.
\end{align}
Applying this relation, along with its $i \leftrightarrow j$ partner, to the two terms on the RHS of \cref{eq:IBP_trivial_top_sector}, makes those two terms cancel and reproduces the zero on the LHS of \cref{eq:IBP_trivial_top_sector}. Thus, this IBP identity can be derived from subsectors only, and therefore does not satisfy the \textit{second} requirement of a \hyperlink{box:magicDef}{magic relation}.

To ensure that the full set of syzygies, and not just the generators, is accounted for, the case $f\neq1$ must be considered. To maintain the divergence free property $f$ must satisfy
\begin{equation}
    0 = \partial_if\, \partial_j \mathcal{G} -\partial_j f \,\partial_i \mathcal{G}
    =
    |J|
    =
    \begin{vmatrix}
    {\partial_i \mathcal{G}} & {\partial_j \mathcal{G}} \\[0.5mm]
    {\partial_i f} & {\partial_j f}
    \end{vmatrix}
    \,.
\end{equation}
This implies that there exists an eigenvector $n$ such that $n \cdot \nabla f = n \cdot \nabla \mathcal{G} = 0$. That is, $n$ is normal to both $\nabla f$ and $\nabla \mathcal{G}$ in the $\{i,j\}$ coordinate space, which implies
\begin{equation}\label{eq:Gfrelation}
    \nabla f \parallel \nabla \mathcal{G}\,.
\end{equation}
This condition places strong constraints on the relation between $f$ and $\mathcal{G}$. In particular given two univariate polynomials $u$, $v$, and a two-variable polynomial $h$ then \cref{eq:Gfrelation} implies that
\begin{equation}\label{eq:f_g_general_form}
    \mathcal{G}(x) = v(h(x_i,x_j))\,, \qquad f(x) = u(h(x_i,x_j))\,,
\end{equation}
where $u$, $v$ and $h$ will in general have coefficients which depend on all variables except $\{x_i,x_j\}$. From \cref{eq:Gfrelation} and the chain rule, for \cref{eq:gcd_condition_2} to be respected, it is necessary that
\begin{equation}
    \frac{\partial v}{\partial h} = \text{const}\,,\qquad v(h) = a +b\,h\,.
\end{equation}
This implies that $f$ is a (polynomial) function in $\mathcal{G}$:
\begin{equation}
   f = u\left(\frac{\mathcal{G}-a}{b}\right) = \sum_{k=0}^{n}c_k\, \mathcal{G}^k\,, \quad \partial_i\, c_k = \partial_j\, c_k = 0\,.
\end{equation}
Substituting this form into eq.~\eqref{eq:LP_Syz_IBP_2} the argument proceeds identically to above for each term in $f$.

\bibliographystyle{JHEP}
\bibliography{biblio.bib}

\end{document}